\documentclass[twocolumn]{aastex631}
\usepackage{esint, multirow}

\shorttitle{The Milky Way Halo Stars toward the Galactic Center}
\shortauthors{Yang et al.}
\graphicspath{{./}{figures/}}
\def\kms{\mathrm{km\,s}^{-1}}
\def\kmskpc{\mathrm{km\,s^{-1}\,kpc^{-1}}}
\def\vphi{v_{\phi}}
\def\feh{{\rm[Fe/H]}}
\def\ldisk{\mathcal{L}_{\rm disk}}
\def\lhalo{\mathcal{L}_{\rm halo}}
\def\pdisk{\mathcal{P}_{\rm disk}}
\def\phalo{\mathcal{P}_{\rm halo}}
\def\nph{n_{\rm ph}}
\def\nsp{n_{\rm sp}}
\def\nuph{\nu_{\rm ph}}
\def\nusp{\nu_{\rm sp}}
\def\rgc{r}
\def\teff{T_{\rm eff}}
\def\logg{{\rm log}(g)}
\def\vr{v_{r}}

\begin{document}
\title{Constructing the Milky Way Stellar Halo in the Galactic Center by Direct Orbit Integration}

\correspondingauthor{Ling Zhu}
\email{lzhu@shao.ac.cn}

\author[0000-0003-1972-0086]{Chengqun Yang}
\affiliation{Shanghai Astronomical Observatory, Chinese Academy of Sciences, 80 Nandan Road, Shanghai 200030, China; 
    \href{mailto:ycq@shao.ac.cn}{ycq@shao.ac.cn}}

\author[0000-0002-8005-0870]{Ling Zhu}
\affiliation{Shanghai Astronomical Observatory, Chinese Academy of Sciences, 80 Nandan Road, Shanghai 200030, China; \href{mailto:ycq@shao.ac.cn}{ycq@shao.ac.cn}}

\author[0000-0002-1584-2281]{Behzad Tahmasebzadeh}
\affiliation{Shanghai Astronomical Observatory, Chinese Academy of Sciences, 80 Nandan Road, Shanghai 200030, China; \href{mailto:ycq@shao.ac.cn}{ycq@shao.ac.cn}}

\author[0000-0002-0642-5689]{Xiang-Xiang Xue}
\affiliation{National Astronomical Observatories, Chinese Academy of Sciences, 20A Datun Road, Chaoyang District, Beijing 100101, P.R.China}

\author[0000-0002-1802-6917]{Chao Liu}
\affiliation{National Astronomical Observatories, Chinese Academy of Sciences, 20A Datun Road, Chaoyang District, Beijing 100101, P.R.China}
\affiliation{School of Astronomy and Space Science, University of Chinese Academy of Sciences, 19A Yuquan Road, Shijingshan District, Beijing 100049, P.R.China}


\begin{abstract} 
The halo stars on highly-radial orbits should inevitably pass the center regions of the Milky Way. 
Under the assumption that the stellar halo is in ``dynamical equilibrium" and is axisymmetric, we integrate the orbits of $\sim 10,000$ halo K giants at $5\leq \rgc \leq 50$ kpc cross-matched from LAMOST DR5 and Gaia DR3. By carefully considering the selection function, we construct the stellar halo 
distribution at the entire regions of $\rgc \leq 50$ kpc.
We find that a double-broken power-law function well describes the stellar halo’s density distribution with shallower slopes in the inner regions and the two breaks at $\rgc=10$ kpc and $\rgc=25$ kpc, respectively. The stellar halo becomes flatter from outer to inner regions but has $q\sim 0.5$ at $\rgc \lesssim 5$ kpc.
The stellar halo becomes isotropic with a slight prograde rotation in the inner 5 kpc, and reaches velocity dispersions of $\sim 250\ \kms$.
We get a weak negative metallicity gradient of $-0.005$ dex kpc$^{-1}$ at $5\leq r \leq 50$ kpc, while there is an excess of relative metal-rich stars with [Fe/H]$>-1$ in the inner 10 kpc.
The halo interlopers at $\rgc \leq 5$ kpc from integration of our sample has a mass of $\sim1.2 \times 10^8 M_{\odot}$ ($\sim 4.7 \times 10^7 M_{\odot}$ at $\feh<-1.5$), which can explain $50-100\%$ of the metal-poor stars with $\feh<-1.5$ directly observed in the Galactic central regions.
\end{abstract}



\section{Introduction} \label{sec:intro}

Our Milky Way (MW) Galaxy is a large spiral galaxy containing a flattened disk, a central bulge/bar, and a halo. 
The most visible part of the MW is its disk, which can be decomposed into a denser thin disk with a wide range of ages \citep{Robin03, Sharma11, Sharma19, Belokurov20, Franchini20} and an older, more diffuse, and metal-poor thick disk \citep{Chiba00, Bensby03, Bensby05, Reddy06, Kilic17}.
The bulge/bar is the most concentrated component and is dominated by old and metal-rich stars \citep{McWilliam94, Clarkson08, koch16, Williams16, Arentsen20b, Arentsen21, Queiroz21, Lucey21}.
The halo surrounding the disk and bulge/bar, is divisible into two broadly overlapping structural components: an inner and an outer halo. The inner halo is an oblate and probably triaxial component \citep{Bell08, Juric08, Robin14, Perez-Villegas17}, and the outer halo is assumed to be much more spherical \citep{Carollo07, Akhter12, Xue15, Pila15, Liu17, Xu18}. The halo is the MW's most spatially extended and kinematically hot component.

The composition of the Galactic center is complicated due to the blending of different structures.
The metallicity-distribution function from red clump stars (RCs) shows that the Galactic center could have five components, including two metal-rich bulge components, one thin disk component, one thick disk component, and a metal-poor halo component \citep{Ness13a}. 
Additionally, the metal-poor ($\feh < 0$) and metal-rich stars ($\feh > 0$) in the bulge have very different velocity dispersion profiles, indicating their different physical origins \citep{Zoccali17, Rojas17}. 
Now it is clear that the metal-rich stars in the Galactic center are the major ingredient of the B/P structure \citep{Williams16, Barbuy18}, and the less metal-rich stars could be part of the thick disk. 

However, the origin of the most metal-poor stars in the Galactic center is still unclear. They could be part of the confined classical bulge stars formed in-situ \citep{Babusiaux10, Hill11, Zoccali14}, or just passing through stars from the overlapping halo \citep{Debattista17}.
There might be a small part of metal-poor halo stars trapped into the bar structure \citep{Perez-Villegas17}.
A significant fraction of the chemically anomalous stars (mostly metal-poor) identified in the inner Galaxy has been demonstrated to be interloper stars that have likely escaped from globular clusters \citep{Schiavon17, Recio-Blanco17, Lucey19, Fernandez-Trincado19a, Fernandez-Trincado19b, Fernandez-Trincado20a, Fernandez-Trincado20b, Horta21}.

It is generally accepted that the halo stars are at least partly from accretion \citep{HW99, BJ05, Johnston08, Cooper10, Font11, Conroy19b}. They have highly-radial orbits over almost the whole galaxy \citep{Bird19}. The halo stars on highly-radial orbits should inevitably pass the Galactic center and contribute part of it. Rather than the stars' current positions $\rgc$, the apocentric distance $r_{\rm apo}$ of the stellar orbits provides us with a more physical definition of the stellar halo. In this sense, halo stars should distribute in both the Galactic center and the outer regions.

The Galactic center is highly obscured and blended with the crowded bulge/bar, and it is hard to identify halo stars directly and evaluate their contribution. With the data release of Gaia \citep{Gaia18}, the full six-dimensional phase-space parameters of stars in a few small fields in the Galactic center are available. Unbound stars on orbits with large apocentric distances are indeed found and are considered as halo interlopers. \citet{Kunder20} found that 25\% of 1389 RR Lyrae stars (RRLs) in the Galactic center are halo interlopers, while \citet{Lucey21} suggested that $\sim 50\%$ of the 523 stars in their sample are halo interlopers, including red giant branch stars (RGBs) and subgiant stars. And \citet{Rix22} found a minor fraction ($<50\%$) of halo interlopers in their large sample of $\sim$ 18000 stars with $\feh<-1.5$ toward the Galactic center.

Beyond the Galactic center, the stellar halo has been studied extensively. For example, the density profile of the stellar halo can be described as a single power law with a varied flattening or a broken power law with constant flattening \citep{Xue15}. The broken position at $\rgc \sim$ 20 kpc, indicates the transition between the ``inner'' and ``outer'' halo, and is probably caused by the early accretion of a massive satellite-like Gaia Sausage Enceladus (GSE) \citep{Deason18, Lancaster19, Myeong18, Myeong19, Massari19, Wu22b}. In \citet{Xu18}, with K giants from the Large Sky Area Multi-Object Fibre Spectroscopic Telescope \citep[LAMOST;][]{Zhao12, Cui12, Luo12} and well-known selection function \citep{Liu17}, they constructed the density profile of the stellar halo up to axial distance $R \sim 35$ kpc, but truncated at $R \lesssim 10$ kpc due to the lack of stars in the inner regions. The profile of the inner 10 kpc was extrapolated inward by assuming a power-law profile.

The last massive merger of the MW is supposed to be happened a long time ago \citep{Helmi18, Belokurov18, Belokurov20}. The smooth MW halo should generally be in ``dynamical equilibrium" and ``phase mixed," excluding these streams created by recent minor mergers. In this sense, the halo interlopers in the Galactic center regions should have their companions on the same orbits but are currently located in the outer regions.
In this work, we fill the missing inner part of the stellar halo by integrating the orbits of a complete halo star sample observed in the outer regions with $5 \le \rgc \le 50$ kpc. We describe the sample selection in Section \ref{sec:data}. We correct the selection function and integrate the orbits of halo stars to obtain the complete spatial, kinematic, and metallicity distribution of the stellar halo in Section \ref{sec:dens}. The halo interlopers in the classic bulge regions are shown in Section \ref{sec:cen}. We discuss the possible impact of disk contamination on our results in Section \ref{sec:Dis} and conclude in Section \ref{sec:sum}.

In this work, we use the Cartesian coordinate that the $x$-axis is positive toward the Galactic center, the $y$-axis is along with the rotation of the disk, and the $z$-axis points toward the North Galactic Pole. The Galactocentric distance is defined as $\rgc =\sqrt{x^2 + y^2 + z^2}$, and we use $R = \sqrt{x^2 + y^2}$ as the distance across the disk plane. We adopt the solar position of $(-8.2,0,0)$ kpc \citep{McMillan17}, which is consistent with the recently determined solar position $R_\odot$ = 8.178 kpc \citep{GC19}. The local standard-of-rest (LSR) velocity is 232.8 km s$^{-1}$ \citep{McMillan17}, and the solar motion is $(+11.1,+12.24,+7.25)$ km s$^{-1}$ \citep{Schonrich10}.

\section{Data} \label{sec:data}

\subsection{LAMOST K-giant stars} \label{subsec:K giants}

\begin{figure*}
\centering
    \begin{tabular}{l}
        \hspace{-2.0cm} \includegraphics[height=.48\textwidth]{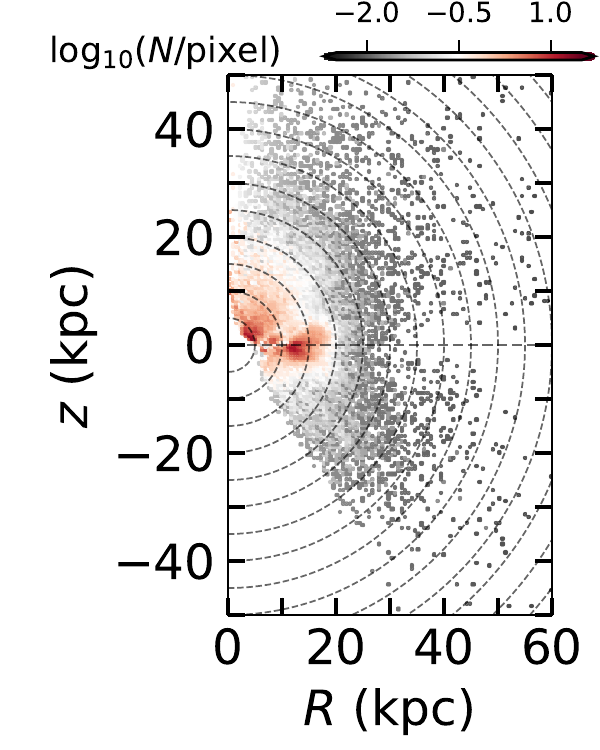}
    \end{tabular}
    \begin{tabular}{r}
        \hspace{-1.0cm} \includegraphics[height=.48\textwidth]{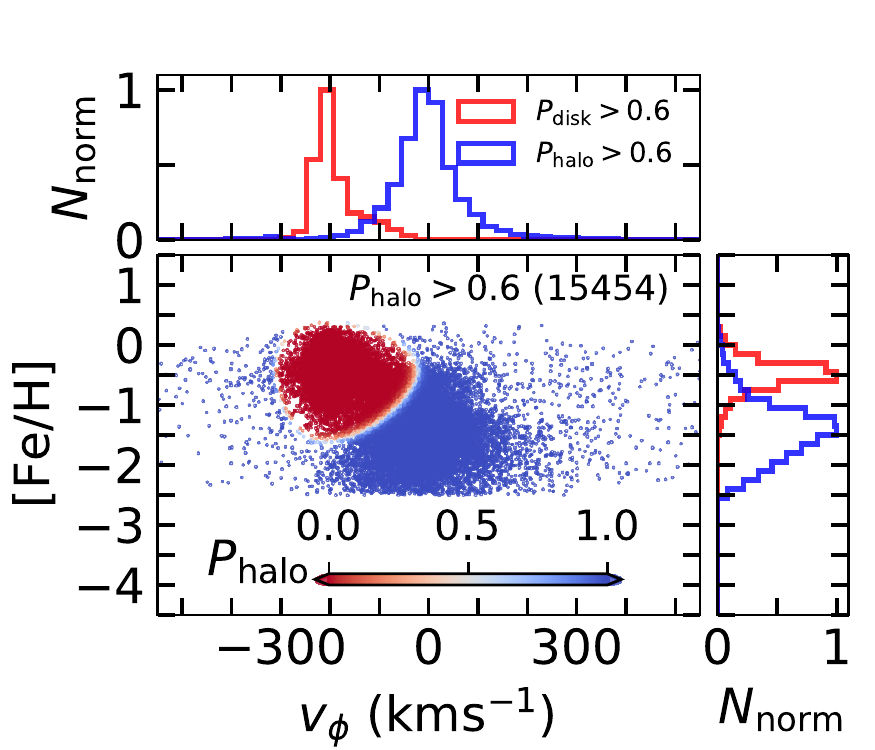}
    \end{tabular}
\caption{
{\bf Left}: the density distribution of 38,457 LAMOST K giants in $(R, z)$ space, the density in the unit of log$_{10}(N$/pixel) is indicated by the color bar. 
{\bf Right}: the distribution of LAMOST K giants in $\vphi$ versus $\feh$ space, color-coded by the probability of being a halo star ($\phalo$). The blue and red histograms in the top (right) subpanel indicate the $\vphi$ ($\feh$) distribution of halo stars ($\phalo>0.6$) and disk stars ($\pdisk>0.6$), respectively.}
\label{xz_vfeh}
\end{figure*}

LAMOST is a large aperture multifiber telescope that observes nearly 4000 stars simultaneously. Its sky coverage is from decl. $\delta = \rm -10^\circ$ to $\delta = \rm 90^\circ$. The targets of LAMOST are selected from various photometric catalogs, e.g., SDSS, Two Micron All Sky Survey \citep[2MASS;][]{Skrutskie06}, Pan-STARRS1 \citep[PS1;][]{Chambers16}, and other catalogs. The first phase, from 2011 to 2018, observed over nine million low-resolution spectra (LRS; $R\sim1,800$) and was published in the fifth data release (DR5). The limiting magnitude of LAMOST LRS is $17.8$ in the SDSS $r$ band, and $\sim$80\% of the spectra have a signal-to-noise ratio (SNR) in the $r$ band larger than 10. The enormous number of spectra provides great help to our understanding of the MW \citep{Liu14, Tian15, Xiang15, Liu17, Xu18, Li19, Tian20, Xiang22}.

From the LAMOST spectra, we obtain the stars' effective temperature $\teff$, surface gravity $\logg$, metallicity $\feh$, and heliocentric radial velocities $\vr$ \citep{Koleva09, Wu11}. The mean errors of $\teff,\ \logg,\ \feh$,\ and\ $\vr$ in the LAMOST DR5 catalog are 120 K, 0.19, 0.11, and 6.7 $\kms$, respectively.

K giants are good tracers for the MW stellar halo with widespread spatial and metallicity distributions \citep{Morrison90, Starkenburg09, Xue15}. We use the sample of K giants cross-matched from LAMOST DR5 and Gaia DR3. The K giants are selected as $\logg< 4$ when $4600 <\teff< 5600$ K or $\logg< 3.5$ when $4000 < \teff< 4600$ K \citep{Liu14}, resulting in $\sim$1.1 million stars in LAMOST DR5. We then cross-match with the Gaia DR3 catalog with a radius of $1\arcsec$, and 99\% of these K giants are cross-matched. We thus obtained the proper motions ($\mu_{\alpha}, \mu_{\delta}$) along the equatorial coordinate $(\alpha, \delta)$ of these stars from Gaia DR3. Additionally, based on the Gaia recommended astrometric quality indicator, the renormalized unit weight error (RUWE) \citep{Lindegren21}, in this work, we only use the Gaia DR3 stars with RUWE $< 1.4$ \citep{Fabricius21}, which means the source is astrometrically well behaved.

Estimating distance for K giants is not easy because of their wide extent in the color-magnitude diagram. The distances of our K giants are determined by a Bayesian method based on color-magnitude relation \citep{Xue14}, with a median relative error of $\sim 13\%$. We first derive the fiducial color-absolute magnitude relations of K giants as a function of metallicity by interpolating from K giants in four-star clusters with different metallicity. Then we obtain the absolute magnitudes of each K-giant by comparing the aforementioned relations with its colors from SDSS and metallicity from LAMOST. The distance of each K-giant is thus obtained by comparing the absolute magnitude with the apparent magnitude from PS1 under the consideration of extinction from \citet{Schlegel98}. The deeper magnitude of PS1 helps detect distant halo stars, but PS1 is saturated at $r_{\rm p1} < 13.5$ \citep{Magnier13}. Some nearby stars are thus excluded due to saturation. 

Among $\sim$1.1 million K giants in LAMOST DR5, we only keep the stars that have PS1 magnitude. The stars kept are thus fainter than the saturation magnitude and brighter than the limiting magnitude of PS1 \citep{Chambers16}, and with magnitude errors less than 0.05, $\sim$600,000 stars are excluded from the sample at this step. In addition, the coverage of color ranges is limited in the fiducial color-magnitude relations (see Figure 5 in \citet{Xue14}), we can only obtain absolute magnitude for stars in the corresponding color range covered by the fiducial with their metallicity, $\sim$240,000 stars are thus excluded. We further select the sample to only keep the K giants above the horizontal branch to prevent contamination from red clump stars, $\sim$210,000 stars are further excluded. After removing duplicate stars within 1$\arcsec$, in the end, we have 38,457 K giants with full 6D phase-space information $(l,\ b,\ d,\ \mu_{\alpha},\ \mu_{\delta},\ \vr)$ and metallicity measured. The spatial distribution of the sample is shown in the left panel of Figure \ref{xz_vfeh}. Most of the stars are distributed at $5<\rgc< 50$ kpc. 

Both in the pre-Gaia and Gaia era, the distance of K giants estimated in this way has been widely used and accepted in many works, including constructing the stellar halo density profile and shape \citep{Xue15, Das16, Xu18, Wu22b}, the stellar halo kinematics \citep{Bhattacharjee14, Deason17, Kafle17, Bird19, Petersen21, Bird21, Erkal21}, the stellar halo metallicity profile and chemical abundance \citep{Xue15, Das16, Battaglia17}, obtaining the Milky Way mass \citep{Huang16, Williams17, Zhai18, Deason21, Bird22}, and identifying halo substructures \citep{Deason14b, Janesh16, Simion18, Yuan19, Zhao20, Li21, Wu22a}. We have a direct comparison of the distance we estimated and the distance obtained by photoastrometric method {\tt StarHorse} \citep{Anders22} in Appendix \ref{app:dist}. Distance obtained in these two methods are consistent with each other in $\sim 90\%$ of stars within 20 kpc, but our method is more efficient in providing accurate distance for stars at $d>20$ kpc.

In the left panel of Figure \ref{xz_vfeh}, we show the number density distribution of the final sample of 38,457 LAMOST K giants in $(R, z)$ space. Coupled with the observational design of the LAMOST survey \citep{Deng12}, more stars were observed in the Northern Galactic hemisphere and the Galactic anticenter.

\subsection{Separation of halo and disk} \label{subsec:sep_halo}

\begin{figure*}
\centering
    \hspace{-1.2cm}
    \includegraphics[width=.5\textwidth]{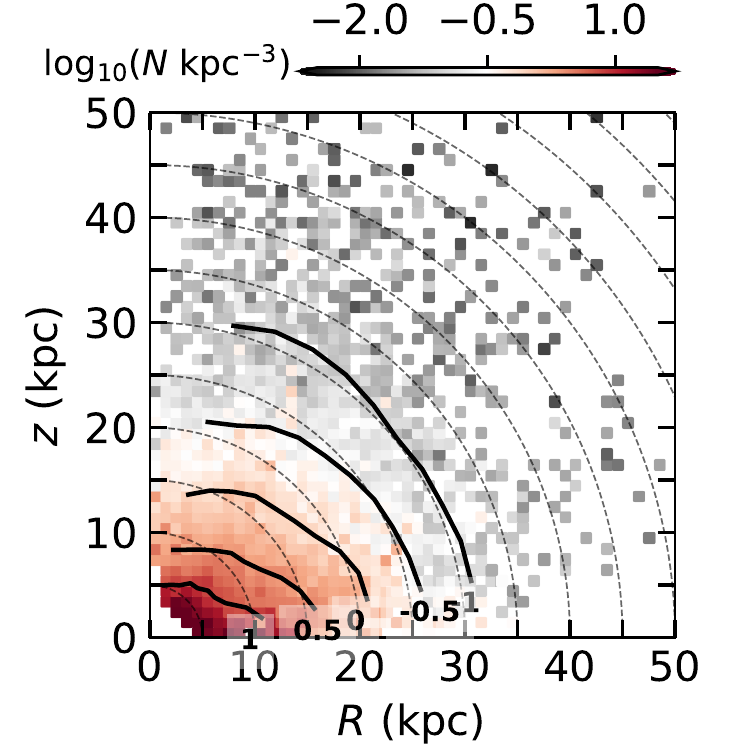}
    \hspace{.0cm}
    \includegraphics[width=.5\textwidth]{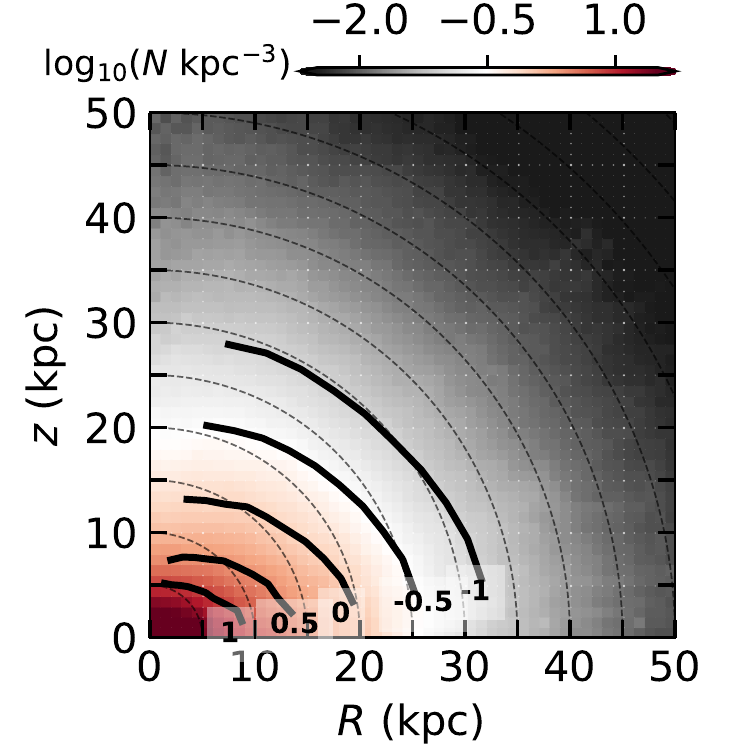}
\caption{
{\bf Left:} the stellar number density distribution of halo K giants in the $(R, z)$ plane, after the correction of the selection effects, the density in the unit of number per ${\rm kpc}^{3}$ is indicated by the color bar. The solid black curves with labeled values are the isodensity contours. The dashed curves are the concentric circles for every 5 kpc. 
{\bf Right:} similar to the left panel, but the number density distribution constructed by particles sampled from the stellar orbits integrated in MW potential model M3. The stars observed at large radius but on highly-radial orbits pass through the Galactic center, thus fill in the density distribution of the very inner regions.
} \label{Den_RZ}
\end{figure*}

Halo and disk stars are mixed in our sample. The halo and disk stars have different spatial, velocity and metallicity distributions. To keep the halo stars that spatially overlapping with the disk, we separate the halo and disk stars not based on their spatial distribution, but on their different distributions in $\feh$ and Galactocentric azimuthal velocity $\vphi$.

We assume the metallicity $\feh$ and velocity $\vphi$ of halo and disk stars follow skewed normal distributions:
\begin{equation}
    f(x) = 2\ \phi(\frac{x-\xi}{\omega})\ \psi(\alpha\ (\frac{x-\xi}{\omega})),
\end{equation}
where $\phi(.)$ and $\psi(.)$ represent the normal probability density function and normal cumulative distribution function. $\xi,\ \omega$, and $\alpha$ are location, scale, and skewness. We obtain the parameters in metallicity distributions of disk $f_{m, \rm disk}(x)$ and halo $f_{m, \rm halo}(x)$ by fitting the clean samples selected with $\vphi <- 180\ \kms$ and $\vphi > 0\ \kms$ for disk and halo, respectively.
Similarly, we obtain the parameters in the $\vphi$ distributions of disk $f_{v, \rm disk}(x)$ and halo $f_{v, \rm halo}(x)$ by fitting the clean samples with $\feh > -0.5$ and $\feh < -1.0$ for disk and halo, respectively. 

Then for a star having $\feh=m$ and $\vphi=v$, we can calculate its probability of being a halo and disk star by combing the metallicity and velocity distributions:
$
\lhalo = f_{m, \rm halo}(m) \times f_{v, \rm halo}(v)
$
and
$
\ldisk = f_{m, \rm disk}(m) \times f_{v, \rm disk}(v)
$. 
Its normalized probability of being halo is $\phalo = \lhalo/(\ldisk + \lhalo)$. 
In the right panel of Figure \ref{xz_vfeh}, we show the distribution of K giants in $\feh$ vs. $\vphi$ color-coded by $\phalo$.
Over 97\% of stars in our sample have $\phalo$ either smaller than 0.1 (high probability of being disk stars) or larger than 0.9 (high probability of being halo stars); disk and halo stars are generally well separated. There are some stars with $\phalo$ in between. We choose stars with $\phalo >$ 0.6 as halo stars in the final sample, which results in 15,454 stars. 

The data is complete in the northern Galactic hemisphere, which covers $\sim 90^\circ$ in the first quadrant of $(R,z)$ space and $\sim 60^\circ$ in the fourth quadrant of $(R,z)$ space, while the data in the South Galactic hemisphere are much less complete. This work assumes the halo is axisymmetric and only uses halo stars in the northern hemisphere (10,651 stars) in the following analysis.

\section{Density distribution of halo K giants} \label{sec:dens}

We take two major steps to construct a full number density distribution of the halo K giants: (1) correction of selection function, (2) orbit integration, by which we will fill the Galactic center lacking of direct observations, taking the assumption that the stellar halo is axisymmetric and in ``dynamical equilibrium.''

\subsection{Correction of selection function} \label{sec:dens:st_dens}

\begin{table*}
\centering
\caption{Model parameters for the MW. 
}  \label{pots}
\footnotesize
\setlength\tabcolsep{6pt}
\begin{tabular}{ccccccccccccc}
\hline
\hline
\colhead{Model} &
\colhead{${M_{\rm halo}   }^{\rm a}$} &
\colhead{${r_{\rm vir}    }^{\rm b}$} &
\colhead{${c              }^{\rm c}$} &
\colhead{${M_{\rm disk}   }^{\rm d}$} &
\colhead{${a_{\rm disk}   }^{\rm e}$} &
\colhead{${b_{\rm disk}   }^{\rm f}$} &
\colhead{${M_{\rm bar}    }^{\rm g}$} &
\colhead{${a_{\rm bar}    }^{\rm h}$} &
\colhead{${p_{\rm bar}    }^{\rm i}$} &
\colhead{${q_{\rm bar}    }^{\rm j}$} &
\colhead{${\phi_{\rm bar} }^{\rm k}$} &
\colhead{${\Omega_b       }^{\rm l}$}
\vspace{-0.5em}
\\
\colhead{} &
\colhead{$\rm{(M_\odot)}$} &
\colhead{(kpc)} &
\colhead{} &
\colhead{$\rm{(M_\odot)}$} &
\colhead{(kpc)} &
\colhead{(kpc)} &
\colhead{$\rm{(M_\odot)}$} &
\colhead{(kpc)} &
\colhead{} &
\colhead{} &
\colhead{($\rm ^o$)} &
\colhead{($\kmskpc$)}
\\
\hline
M1 & $0.8\times 10^{12}$ & $251.8$ & 20.6 & $8.31\times 10^{10}$ & 3.0 & 0.28 & 
$1.2\times 10^{10}$ & 3.5 & 0.44 & 0.31 & $-25$ & $-40$ \\
M2 & $1.0\times 10^{12}$ & $271.2$ & 22.8 & $8.08\times 10^{10}$ & 3.0 & 0.28 & 
$1.2\times 10^{10}$ & 3.5 & 0.44 & 0.31 & $-22$ & $-40$  \\
M3 & $1.0\times 10^{12}$ & $271.2$ & 22.8 & $8.08\times 10^{10}$ & 3.0 & 0.28 & 
$1.2\times 10^{10}$ & 3.5 & 0.44 & 0.31 & $-25$ & $-40$  \\
M4 & $1.0\times 10^{12}$ & $271.2$ & 22.8 & $8.08\times 10^{10}$ & 3.0 & 0.28 & 
$1.2\times 10^{10}$ & 3.5 & 0.44 & 0.31 & $-28$ & $-40$  \\
M5 & $1.0\times 10^{12}$ & $271.2$ & 22.8 & $8.08\times 10^{10}$ & 3.0 & 0.28 & 
$1.2\times 10^{10}$ & 3.5 & 0.44 & 0.31 & $-25$ & $-35$  \\
M6 & $1.0\times 10^{12}$ & $271.2$ & 22.8 & $8.08\times 10^{10}$ & 3.0 & 0.28 & 
$1.2\times 10^{10}$ & 3.5 & 0.44 & 0.31 & $-25$ & $-45$  \\
M7 & $1.3\times 10^{12}$ & $296.0$ & 25.7 & $7.79\times 10^{10}$ & 3.0 & 0.28 & 
$1.2\times 10^{10}$ & 3.5 & 0.44 & 0.31 & $-25$ & $-40$  \\
\hline
{\bf Notes.} \\
\multicolumn{13}{l}{\textsuperscript{a}{ Mass of dark matter halo contained within the virial radius.}}\\
\multicolumn{13}{l}{\textsuperscript{b}{ The virial radius of the dark matter halo.}}\\
\multicolumn{13}{l}{\textsuperscript{c}{ The concentration parameter of the dark matter halos.}}\\
\multicolumn{13}{l}{\textsuperscript{d}{ Mass of the exponential disk.}}\\
\multicolumn{13}{l}{\textsuperscript{e}{ The scale length of the disks.}}\\
\multicolumn{13}{l}{\textsuperscript{f}{ The scale height of the disks.}}\\
\multicolumn{13}{l}{\textsuperscript{g}{ Mass of the bar.}}\\
\multicolumn{13}{l}{\textsuperscript{h}{ The half-length of the bar.}}\\
\multicolumn{13}{l}{\textsuperscript{i}{ The intermediate-to-major axis ratio of the bar.}}\\
\multicolumn{13}{l}{\textsuperscript{j}{ The minor-to-major axis ratio of the bar.}}\\
\multicolumn{13}{l}{\textsuperscript{k}{ Bar angle between the major-axis of the bar and the line-of-sight to the MW center.}}\\
\multicolumn{13}{l}{\textsuperscript{l}{ The pattern speed of the bar.}}\\
\end{tabular}
\end{table*}

Due to complicated target selection and observational bias, it is challenging to sample a large region of the sky to completeness for a spectroscopic survey. The correction of selection effects of a spectroscopic survey compared to a complete photometric survey is needed before we can study the real density profile of the stellar halo.

We correct the selection effects of the LAMOST survey following the method provided in \citet{Liu17}, where the photometric data is supposed to be completed within its limiting magnitude. In a small color-magnitude plane $(c, m)$ along each line of sight $(l,b)$, the photometric density $\nuph$ is equal to the spectroscopic density $\nusp$ times its selection function $S^{-1}$
\begin{equation}
    \nuph(d|c, m, l, b) = \nusp(d|c, m, l, b) S^{-1}(c, m, l, b),
\end{equation}

The spectroscopic density $\nusp(d|c, m, l, b)$ is calculated through the kernel density estimation (KDE) method:
\begin{equation}
    \nusp(d|c, m, l, b) = \frac{1}{\Omega d^2} \sum_{i}^{\nusp(c,m,l,b)} p_i(d),
\end{equation}
where $p_i(d)$ is the probability density function of distance for the $i_{\rm th}$ star, and $\Omega d^2$ is the volume element between $d$ and $d + \Delta d$.

The selection function $S^{-1}(c, m, l, b)$ is evaluated from
\begin{equation}
    S^{-1}(c, m, l, b) = \frac{\nsp (c,m,l,b)}{\nph (c,m,l,b)},
\end{equation}
where $\nsp (c,m,l,b)$ and $\nph (c,m,l,b)$ are the number of spectroscopic and photometric stars within $(c,m,l,b)$.

Then, the stellar density along a given line of sight is: 
\begin{equation}
    \nuph(d|l, b) = \iint \nusp(d|c, m, l, b) S^{-1}(c, m, l, b){\rm d}c{\rm d}m.
\end{equation}

The left panel of Figure \ref{Den_RZ} presents the 2D density distribution of the stellar halo in the $(R, z)$ plane after correcting the selection effects. The bin size in this figure is $1.0 \times 1.0$ kpc. Since no halo stars are directly detected in the inner $\sim3$ kpc, the central region is missing. The stellar halo tends to be more oblate in the inner region than in the outer region.

With the 2D density distribution map constructed after correcting the selection function, we calculate the photometric weight $\omega_i$ for each star $i$ based on its position $(R_i, Z_i)$ in turn. We consider that all the stars in a $1.0 \times 1.0$ kpc bin have the same weights, and the weight of any star in this bin is:
\begin{equation}
    \omega_i = \frac{\bar \nuph \times V}{\nsp}, 
\label{eq:weight}
\end{equation}
where $\bar \nuph$ is the mean photometric density of the bin that star $i$ is located in, $V$ is the bin's volume, and $\nsp$ is the number of stars we have in this bin.

\subsection{MW Potential} \label{subsec:pot}

\begin{figure*}
\centering
    \hspace{-0.0 cm} \includegraphics[width=1.0\textwidth]{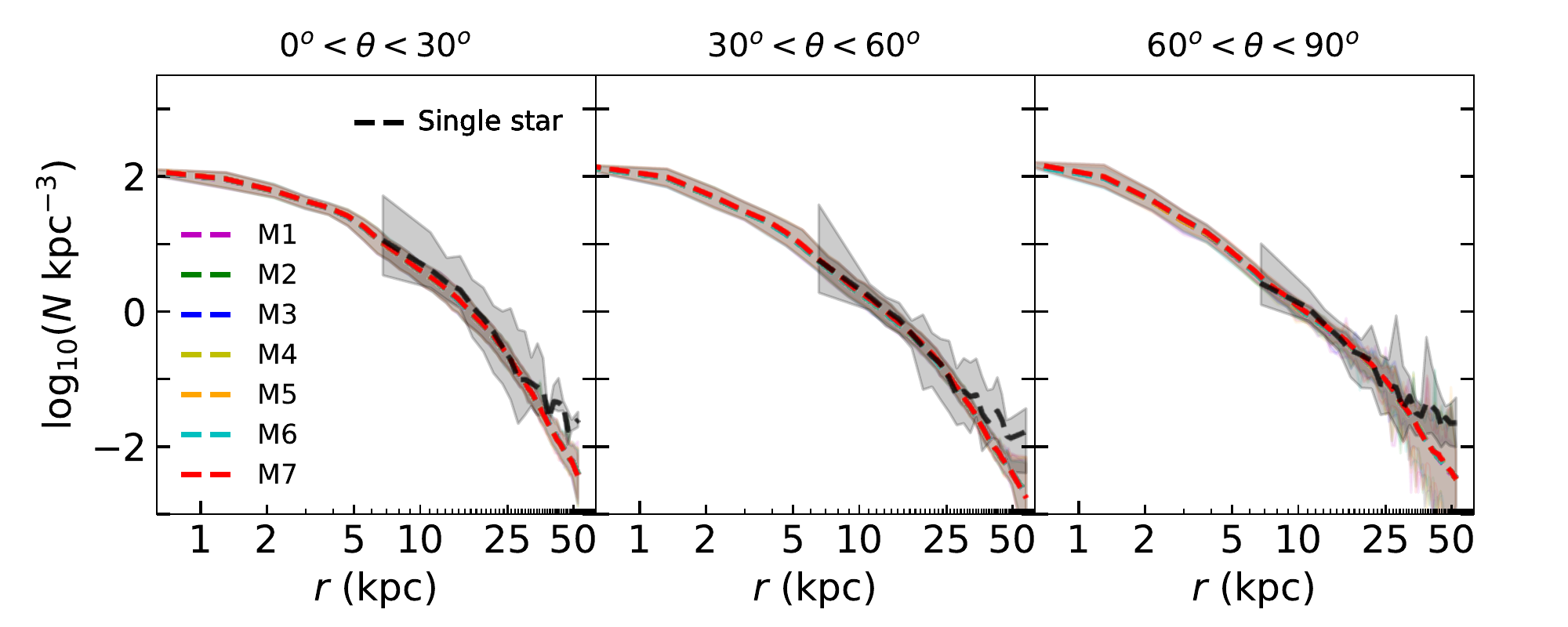}
\caption{The density profiles from halo K giants (black) and the orbital particles (colored) as a function of $\rgc$. From left to right, we show the averaged density profile (the dashed curve) across the elevation angles of $0^{\rm o}<\theta<30^{\rm o}$, $30^{\rm o}<\theta<60^{\rm o}$, $60^{\rm o}<\theta<90^{\rm o}$, respectively. The shadowed regions mark the minimum and maximum of the curve across the region of $\theta$. The orbital particles are integrated in seven different MW models of Table \ref{pots} (shown in different colors), but they are entirely the same and consistent with the stellar density profiles in the overlapping regions. 
} \label{rnu}
\end{figure*}

We integrate the orbits of the halo stars in fixed MW potentials, combining a dark matter halo, a disk, and a bar:
\begin{equation}
    \Phi_{\rm tot} = \Phi_{\rm halo} + \Phi_{\rm disk} + \Phi_{\rm bar}.
\end{equation}

We adopt an NFW halo \citep{Navarro97} with
\begin{equation}
    \Phi_{\rm halo} = - \frac{GM_{\rm halo}}{r}{\rm ln}(1+\frac{rc}{r_{\rm vir}}).
\end{equation}
The halo mass $M_{\rm halo}$ and virial radius $r_{\rm vir}$ are related as:
\begin{equation}
    M_{\rm halo} = \frac{4\pi}{3} \rho_{\rm cr} \Omega_{\rm m} \delta_{\rm th} r_{\rm vir}^3,
\end{equation}
where $\rho_{\rm cr} = 3H^2/8 \pi G$ is the critical density of the universe, 
$\Omega_{\rm m}$ is the contribution of matter to the critical density, and $\delta_{\rm th}$ is the critical over-density within the virial radius. Here we adopt $H = 65\ \kms {\rm M pc^{-1}}$, $\delta_{\rm th} = 340$, and $\Omega_{\rm m} = 0.3$.

There are two free parameters in the NFW halo: halo mass $M_{\rm halo}$ and concentration $c$. 
We adopt three different dark matter halo models with $M_{\rm halo} = 0.8\times 10^{12}\ M_\odot$, $1.0\times 10^{12}\ M_\odot$, and $1.3\times 10^{12}\ M_\odot$, corresponding to the $1\sigma$ lower limit, best-fit value, and $1\sigma$ upper limit of the MW halo mass obtained in \citet{Xue08}, respectively. 
We fix the concentration $c$ according to its correlation with the halo mass:
\begin{equation}
    {\rm log}_{10} c = 1.075 - 0.12({\rm log}_{10}M_{\rm halo} - 12).
\end{equation}

We use an exponential disk \citep{Miyamoto75} with
\begin{equation}
    \Phi_{\rm disk} = - \frac{GM_{\rm disk}}{\sqrt{R^2+(a_{\rm disk}+\sqrt{z^2+b_{\rm disk}^2})^2}}, 
\end{equation}
where disk mass $M_{\rm disk}$, scale length $a_{\rm disk}$, and scale height $b_{\rm disk}$ are free parameters. 
We choose a corresponding disk mass for each fixed halo to keep the circular velocity of $232.8\ \kms$ unchanged at the solar radius of 8.2 kpc. 
The disk masses $M_{\rm disk}$ are chosen to be $8.31 \times 10^{10}\ M_\odot$, $8.08 \times 10^{10}\ M_\odot$, and $7.79 \times 10^{10}\ M_\odot$ for the three halo models. We fix the disk scale radius $a_{\rm disk} = 3.0$ kpc, and scale height $b_{\rm disk}=0.28$ kpc \citep{Bovy15}.

We use a triaxial Ferrers bar model \citep{Ferrers1877, Pfenniger84}, with the density in the form of
\begin{equation}
    \rho = \frac{105M_{\rm bar}}{32 \pi\ p_{\rm bar}\ q_{\rm bar}\ a_{\rm bar}^3}[1 - (\frac{\tilde{r}}{a_{\rm bar}})^2]^2,
\end{equation}
where $\tilde{r} = \sqrt{x_{\rm bar}^2 + (y_{\rm bar}/p_{\rm bar})^2 + (z_{\rm bar}/q_{\rm bar})^2}$.
We fix the parameters as bar mass $M_{\rm bar} = 1.2 \times 10^{10}\ M_\odot$, bar length $a_{\rm bar}$ = 3.5 kpc, 
the axis ratios $p_{\rm bar} = 0.44$ and $q_{\rm bar} = 0.31$ \citep{Queiroz21}.

The MW bar is not aligned with line of sight to the MW center but has a nonzeros bar angle with
\begin{equation}
    \begin{array}{l}
	x_{\rm bar} = x\ {\rm cos(\phi_{bar})} + y\  {\rm sin(\phi_{\rm bar})}, \\
	y_{\rm bar} = - x\ {\rm sin(\phi_{bar})} + y\ {\rm cos(\phi_{\rm bar})}, \\
	z_{\rm bar} = z.
    \end{array}
\end{equation}
We adopt three options of the bar angle with $\phi_{\rm bar} = -22^{\rm o}, -25^{\rm o}, -28^{\rm o}$, corresponding to the $1\sigma$ lower limit, central value, and $1\sigma$ upper limit from \cite{Clarke22}.

We consider the figure rotation of the galaxy and adopt the bar's pattern speed of $\Omega_b = - 40\ \kmskpc$ \citep{Portail17a, Bovy19, Sanders19} as the default value. Considering the uncertainty on bar pattern speed, we add the two cases with $\Omega_b = -35\ \kmskpc$ and $-45\ \kmskpc$ in the midsized halo. We have seven MW potential models named M1 to M7, with their parameters shown in Table \ref{pots}. The results based on M3 will be used for illustration throughout the paper.

\subsection{Orbit integration} \label{subsec:orb_int}

In each of the seven potential models, we integrate the orbits in the rotating frame with pattern speed $\Omega_b$, the equation of motion can be written as follows:
\begin{equation}
    \begin{array}{ll}
        \vspace{0.1cm}  
            \dot{x}=v_{x}+\Omega_b y, & 
        \vspace{0.1cm} 
            \dot{v}_{x}=-\frac{\partial \Phi}{\partial x}+\Omega_b v_{y}, \\
            \dot{y}=v_{y}-\Omega_b x, & 
            \dot{v}_{y}=-\frac{\partial \Phi}{\partial y}-\Omega_b v_{x}, \\
            \dot{z}=v_{z}, & 
            \dot{v}_{z}=-\frac{\partial \Phi}{\partial z},
    \end{array}
\end{equation}
where $x,y,z,\dot{x},\dot{y},\dot{z}$ are the position and velocity in the rotating frame, and $v_x, v_y, v_z$ are the velocity in the inertial frame but in the coordinate instantaneously aligned with the rotating bar.

The orbits are integrated with the python package AGAMA \citep{Vasiliev19a}. We initialize each star with the 6D position-velocity information from observations and consider Monte Carlo variations with observation uncertainties. For each star, we perturb the distance, proper motions, and radial velocity nine times around their central values by adding a random value following a Gaussian distribution dispersed with uncertainties of the observables. We thus integrate 10 orbits for each star, with the central value of observables and the nine perturbed values. For each orbit, we integrate 10 Gyr and store 1000 particles at equal time intervals. The positions $x,y,z$ in the corotating frame and velocities $v_x, v_y, v_z$ in the inertial frame are stored for each particle.


\subsection{The density distribution} \label{subsec:dens}

\begin{figure}
\centering
    \hspace{-0.5 cm} \includegraphics[width=0.48\textwidth]{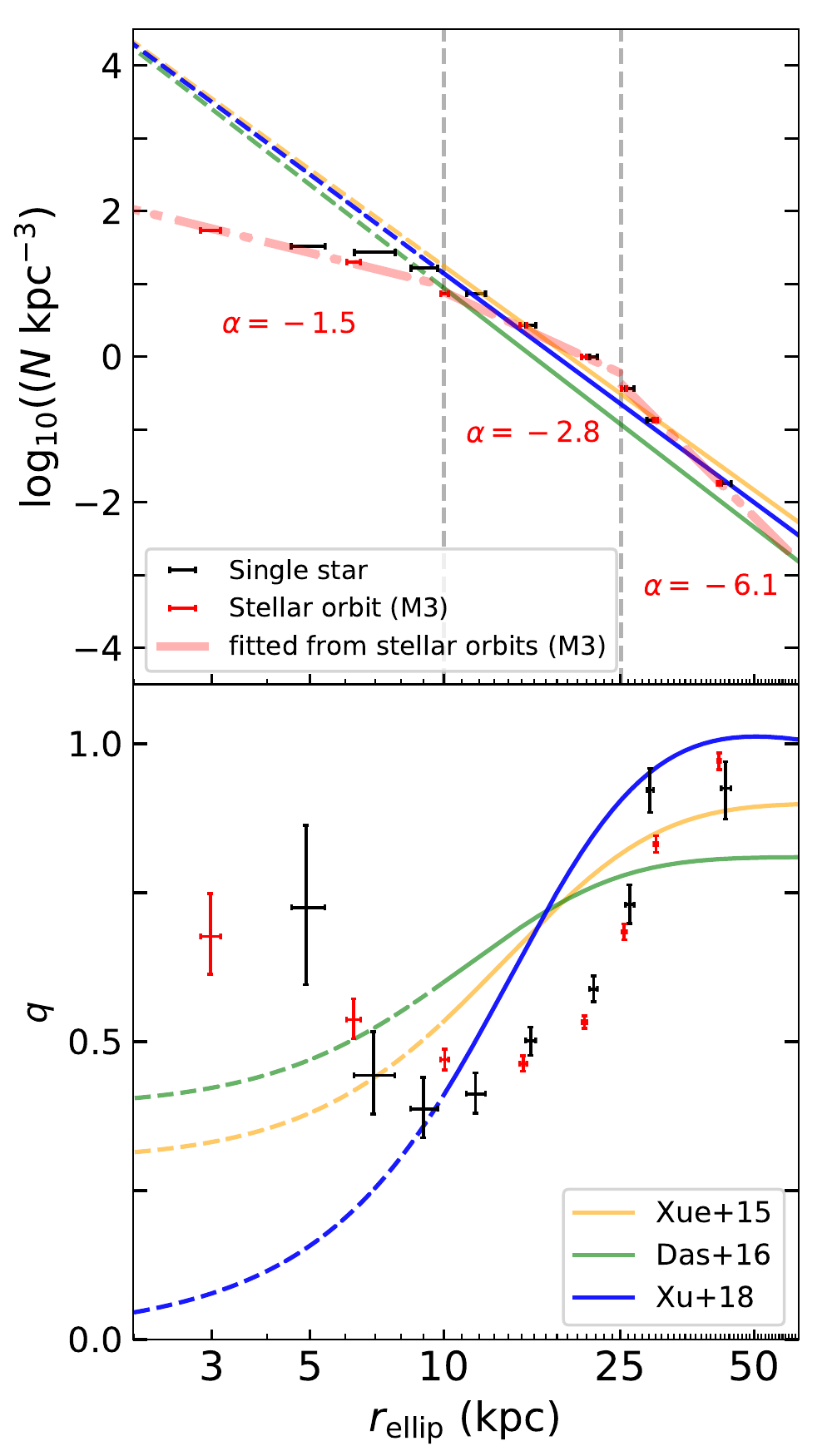}
\caption{
{\bf Top}: the density profile of the stellar halo along the semi-major axis $r_{\rm ellip}$, the black bars indicate the results fitted from single stars, and the red bars are the results fitted from the orbital particles integrated in M3.
The errors are derived from the MCMC process but are very small. The red dashed-dotted lines indicate the double-broken power-law fitting to the red bars, with the power-law coefficients $\alpha$ labeled. The colored lines indicate the density profiles from previous results of \citet{Xue15} (yellow), \citet{Das16} (green), and \citet{Xu18} (blue), and the dashed lines are the models' inward extrapolation.
{\bf Bottom:} the flattening $q$ of the stellar halo along the semi-major axis, symbols are the same as the top panel. The stellar halo becomes flattered from $r\gtrsim 30$ kpc to the inner regions, our results indicate that it keeps $q\gtrsim0.5$ in the inner 5 kpc, unlike the extrapolations of the models from previous work (dashed lines).
} \label{ma_rq_nu}
\end{figure}

\begin{figure*}
\centering
    \hspace{-0.0 cm} \includegraphics[width=1.0\textwidth]{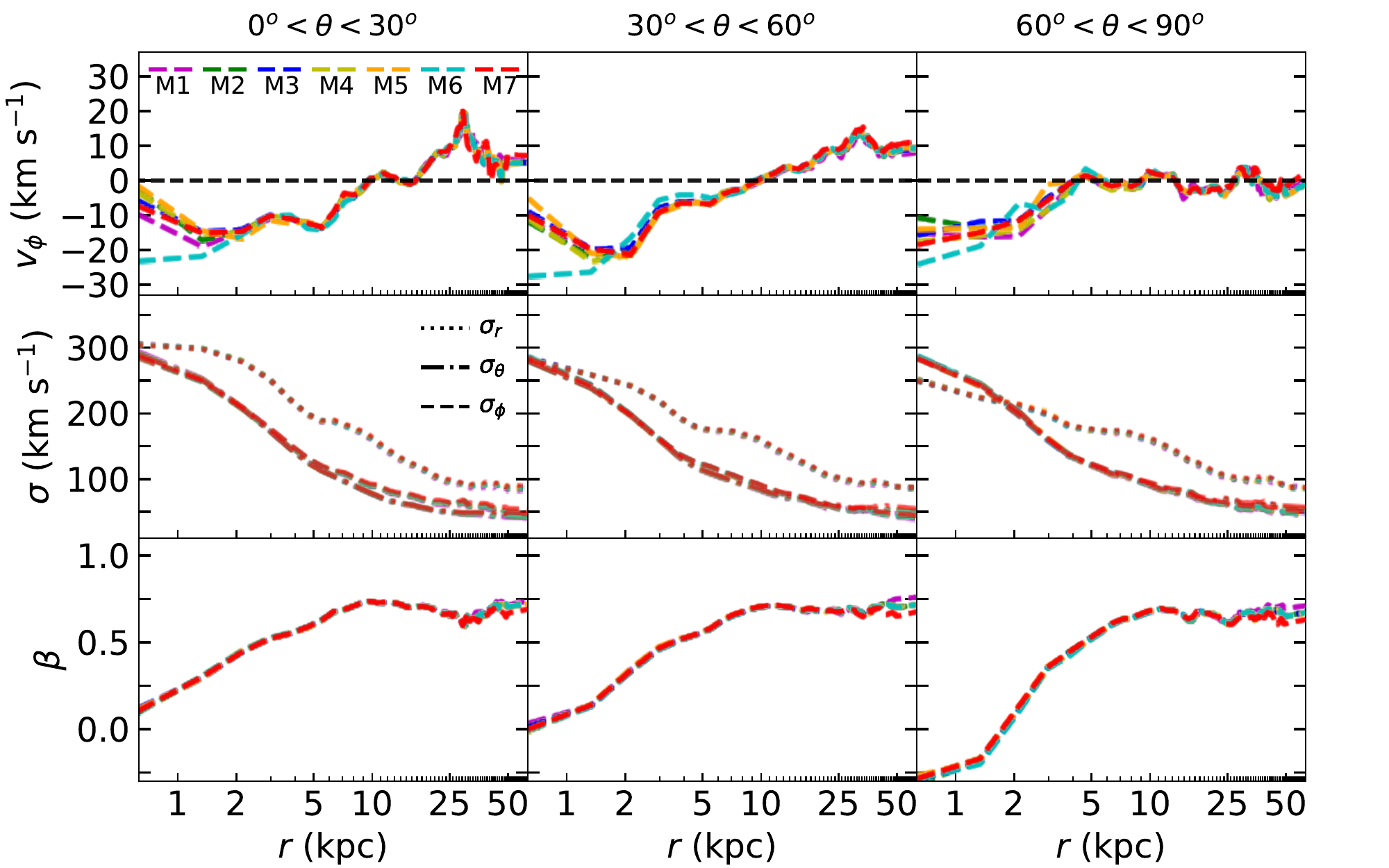}
\caption{Kinematics of the MW stellar halo derived from the orbital particles integrated in the seven MW models and shown in different colors. The panels from left to right indicate the kinematics of the stellar halo from different elevation angular $\theta$. 
{\bf Top}: the mean Galactocentric azimuthal velocity $\vphi$ varies with $\rgc$, and the colored dashed lines shown as M1 to M7 correspond to the seven MW potential models, M5 in orange is the one with the lowest pattern speed of $-35\ \kmskpc$, while M6 in cyan is the one with the highest pattern speed of $-45\ \kmskpc$. The horizontal dashed line indicates zero velocity.
{\bf Middle}: the velocity dispersions in spherical coordinates ($\sigma_r$, $\sigma_{\theta}$, $\sigma_{\phi}$) as a function of $\rgc$. The velocity dispersions are indicated by different line styles, as shown in the legend. 
{\bf Bottom}: the velocity anisotropy parameter $\beta$ as a function of $\rgc$. 
The velocity dispersion and velocity anisotropy profiles obtained from the orbital particles integrated in the seven different MW models are identical. While the bar pattern speed affects the rotation of the halo stars in the inner $\sim 5$ kpc.
} \label{rvl}
\end{figure*}

We construct the full density distribution of the stellar halo using particles from the integrated stellar orbits. Each particle is weighted with the photometric weight $\omega_i$ calculated from Equation (\ref{eq:weight}) and normalized by a factor of 10,000. 
In Figure \ref{Den_RZ}, we show the density distribution constructed directly from single stars in the left panel, and that constructed by particles from orbits integrated in model M3 in the right panel.
The isodensity curves in the two panels are well matched in the overlapped regions. While with particles from the integrated orbits, we can construct stellar halo density maps extending to the very center, thus fill the missing regions of the stellar halo.

We extract the density profiles across different elevation angles with $0^{\rm o}<\theta<30^{\rm o}$, $30^{\rm o}<\theta<60^{\rm o}$, $60^{\rm o}<\theta<90^{\rm o}$, and compare that constructed from single stars and orbital particles integrated in the seven different MW models.
As shown in Figure \ref{rnu}, there is no difference in the density profile of the seven MW models, which are shown in different colors but totally overlapping. In overlapping radial regions (5 $<\rgc<$ 50 kpc), the density profiles constructed from the orbital particles are consistent with that from single stars. It indicates that our assumption of ``in dynamic equilibrium" is reasonable for the stellar halo in general.

We further quantify the surface density maps by fitting the isodensity contours with the following equation
 \citep{Xu18}: 
\begin{equation}
    \rgc = r_{\rm ellip} q\sqrt{\frac{1}{q^2-(q^2-1)sin^2(\theta)}},
\end{equation}
where $r_{\rm ellip}$ is the semi-major axis, and $q$ is the flattening. We separately fit iso-density contours to the surface density map constructed by single stars and the orbital particles integrated in the MW potential M3. Density distributions from other MW potentials are almost identical to M3. Thus, we do not show them here.

The fitted surface density and flattening $q$ as a function of the semi-major axis $r_{\rm ellip}$ are shown in Figure \ref{ma_rq_nu}. As expected, the density and flattening profiles constructed from single stars and that from orbital particles are consistent with each other in the overlapping regions (5 $<r_{\rm ellip}<$ 50 kpc), while the orbital particles allow us to directly obtain surface density and flattening in the inner 5 kpc.

The density profile along the major axis constructed from the orbital particles can be well fitted with a ``double-broken" power-law 
model, with the two breaks at $r_{\rm ellip} \sim 10$ kpc and $r_{\rm ellip} \sim 25$ kpc, the power-law coefficients are $\alpha = -1.5, -2.8, -6.1$ at $r_{\rm ellip} < 10$ kpc, $10< r_{\rm ellip}< 25$ kpc, and $r_{\rm ellip} > 25$ kpc, respectively. This density profile is similar to the double-broken simulation of GSE proposed by \citet{Naidu21}, where the breaks are motivated by the location of the apocenters of GSE. The two breaks of the GSE simulation are located in $\rgc \sim$15 kpc and $\rgc \sim$30 kpc, respectively, with the corresponding power-law coefficients of $\alpha$ = $-1.1, -3.3$, and $-7.0$. This similarity implies that GSE members might be the dominant stars in the inner halo.

In the bottom panel, we show that the stellar halo is near-spherical with $q\sim0.8$ at $r_{\rm ellip}>30$ kpc and it becomes flattered in the inner regions until $r_{\rm ellip}\sim 10$ kpc, generally consistent with previous results \citep{Xue15, Das16, Xu18}. The extrapolations of the flattening profile from the aforementioned works indicate that the halo might become even flatter in the inner 10 kpc. 
However, we found that the halo does not become flatter in the very inner regions, it has $q\gtrsim 0.5$ at $r_{\rm ellip}<5$ kpc, this is consistent with recent dynamical models based on RRLs \citep{Li22} and globular clusters \citep{Vasiliev19b, Wang22}.

\subsection{The kinematic distribution} \label{subsec:kine}

\begin{figure*}
    \hspace{-1.0 cm} \includegraphics[width=0.52\textwidth]{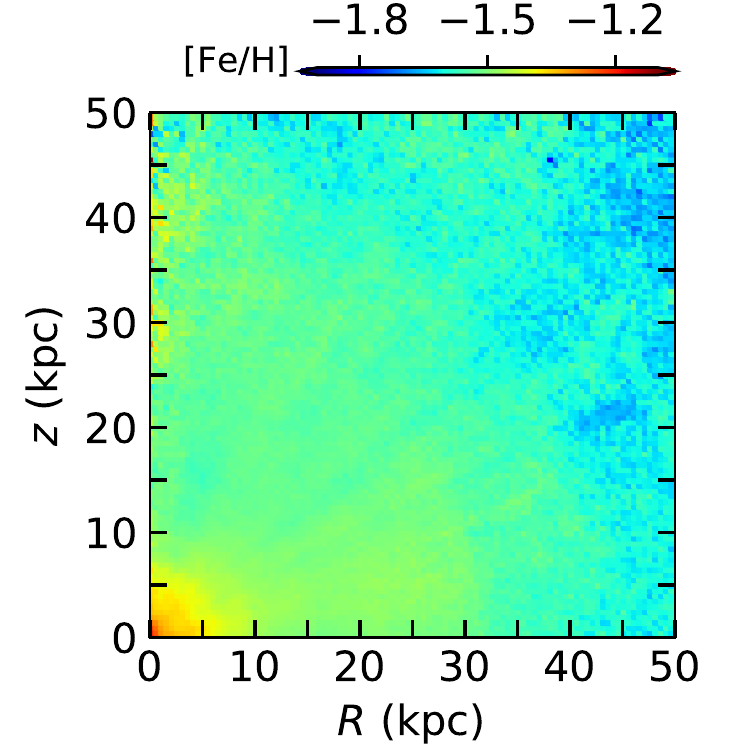}
    \hspace{ 0.0 cm} \includegraphics[width=0.52\textwidth]{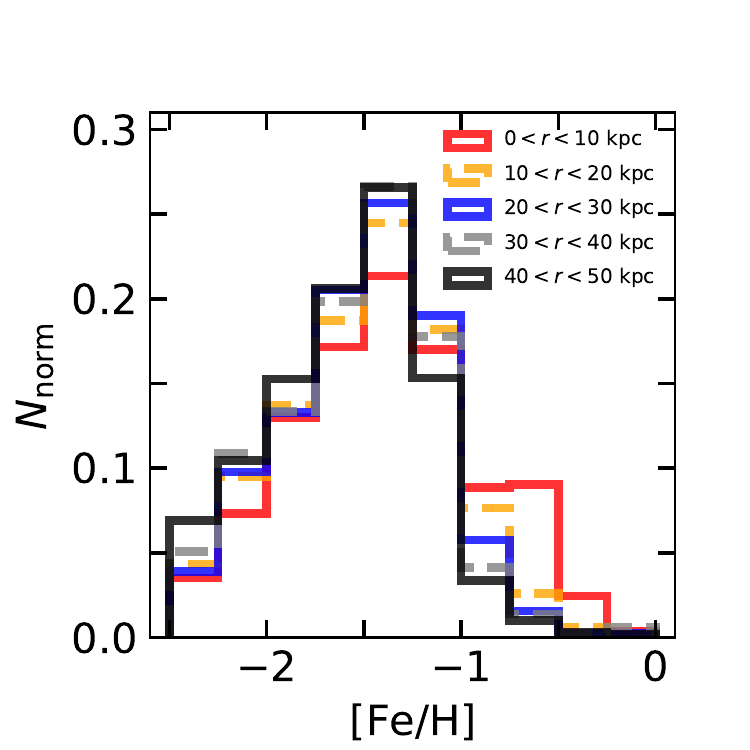}
\caption{The metallicity distribution of stellar halo constructed by the orbital particles integrated in the Milky Way model M3.
{\bf Left}: the metallicity map of stellar halo in $(R,\ z)$ space. The inner region is slightly more metal rich than the outer region. 
{\bf Right}: the metallicity distribution function from different slices of $\rgc$, which are labeled in the upper right corner. The distributions are similar in all bins peaking at $\feh = -1.3$ to $-1.4$. While there is an excess of relative metal-rich stars in the bin with $\rgc < 10$ kpc.
} \label{met}
\end{figure*}

We calculate some basic kinematic properties of the halo with the orbital particles, including the mean Galactocentric azimuthal velocity $\vphi$, velocity dispersions ($\sigma_r$, $\sigma_{\theta}$, $\sigma_{\phi}$), and the velocity anisotropy parameter $\beta = 1 - \frac{1}{2}(\sigma_{\theta}^2 + \sigma_{\phi})/\sigma_{r}^2$, as a function of Galactocentric radius $\rgc$. 
The results are shown in Figure \ref{rvl}, three columns from left to right are the profiles along different elevation angles $\theta$. The results from the orbital particles integrated in the seven different MW models are shown in different colors, results from the seven models are almost identical, except that $\vphi$ in the inner $5$ kpc are different for models with different pattern speeds.

The stellar halo has a mild retrograde rotation of $\sim 10\,\kms$ at the outer region, and a slow prograde rotation of $0$ to $-10\,\kms$ in the inner $\sim$ 5 kpc. 
A similar retrograde rotation of 10 $\kms$ in the outer regions has been seen in many previous works \citep{Beers12, Zuo17, Kafle17, Helmi18, Wegg19}, which could be explained by either a limited number of large merge fragments at these radii \citep{Koppelman18}, or a single large merger with retrograde orbits \citep{Belokurov18, Helmi18}.
While in the inner regions, prograde rotation is also reported in previous works \citep{Perez-Villegas17, Tian19, Wegg19, Vasiliev19b}, but with a much higher velocity of $-30$ to $-80$ $\kms$. The rotation in the inner halo could be
partially caused by disk contamination \citep{Wegg19, Vasiliev19b, Zinn85}.
While in our sample, disk stars have been carefully removed, combining kinematics and metallicity (Figure \ref{xz_vfeh}), which could be a reason for the weaker rotation we obtained.

The velocity dispersion profiles of the halo decrease significantly as a function of radius. The dispersion profiles we obtained at outer regions of the halo are generally consistent with previous studies.
We obtain velocity dispersions of 200-300 $\kms$ for the halo stars at the inner 5 kpc, which is generally consistent with the dispersion of $\sim 200\ \kms$ found for the most metal-poor stars in the Galactic center \citep{Arentsen20a}, but higher than that of 150 $\kms$ based on RRLs \citep{Wegg19} and 100 $\kms$ based on globular cluster \citep{Vasiliev19b, Wang22}. Halo stars selected in our way are guaranteed with apocentric distances larger than 5 kpc, and clean of contamination from bulge stars, which might be the reason for our higher dispersions.


The halo is highly-radially anisotropic with $\beta \sim$ 0.8 in the outer regions, consistent with many previous studies \citep[e.g.,][]{Bird19, Bird21}, while it becomes less radial anisotropic in the inner regions, and almost isotropic with 
 $\beta \sim$ 0 at $\rgc\sim$ 1 kpc, generally consistent with previous studies \citep{Vasiliev19b, Wegg19, Wang22,Li22}.


\subsection{The metallicity distribution} \label{subsec:met}
We obtain the metallicity distribution of stellar halo in the full regions by considering that particles extracted from the same orbit have the same metallicity as the star initialized it. We illustrate the results with orbital particles integrated in the MW potential M3, results from other potential models are identical.

In the left panel of Figure \ref{met}, we present the 2D metallicity distribution map in $(R,\ z)$ space. The stellar halo exhibits a negative metallicity gradient along the radius, steeper at the central 5 kpc with the slope of $-0.029$ dex kpc$^{-1}$, and becomes shallower at larger radii with the slope of $\sim -0.005$ dex kpc$^{-1}$, generally consistent with \citet{Xue15} and \citet{Das16}. Note that we do not impose any hard cut in metallicity for selecting halo stars and this metallicity gradient is not likely to be caused by selection bias.


In the right panel of Figure \ref{met}, we show the metallicity distribution of stars in different bins of $\rgc$. The distributions are similar in all bins with $\rgc > 10$ kpc peaking at $\feh = -1.3$ to $-1.4$. The metallicities peaking values are slightly different from previous studies \citep[e.g.,][]{Horta21, Xue15}, which are likely caused by different ways of selecting halo stars. 
There is an excess of relatively metal-rich stars with $\feh >-1$ in the bin with $\rgc < 10$ kpc, which could be caused by contamination of the ``Splash" structure born in-situ in the disk and being heated to halo-like orbits by an ancient merger \citep{Belokurov20}. We will discuss the effects of Splash stars in Section~\ref{sec:Dis}.

\section{The contribution of Halo Stars in the Galactic center} \label{sec:cen}

After excluding the disk, we have taken the definition of halo stars with apocentric distance $r_{\rm apo} > 5$ kpc and bulge stars with  $r_{\rm apo} < 5$ kpc. Halo stars could be either found at $\rgc > 5$ kpc, or at $\rgc < 5$ kpc but dynamically not bounded. The former is the classic halo stars with complete observations over large areas, as included in our sample, the latter is the so-called ``halo interlopers", which is hard to observe completely in the Galactic center due to dust obscuration \citep{Lucey21, Kunder20, Rix22}. However, halo interlopers should always have companions on the same orbits but are currently located in the outer regions, if the system is in dynamical equilibrium and well phase mixed. With orbits of all halo stars at $\rgc > 5$ kpc integrated, here we evaluate the density of halo interlopers expected to be found in the Galactic center.

\subsection{The total luminosity of the stellar halo} \label{subsec:norm_db}
We take two steps to obtain the total luminosity of the stellar halo. First, we calculate the number of halo K giants $N_{\rm KG}$ from the weighted orbital particles. Then we convert the K giants number to halo luminosity by using the ratio of luminosity to K giants number $L_{\rm halo}/N_{\rm KG}$ derived from isochrones.

\subsubsection{The number of halo K giants} 
We obtained a final sample of 38,457 K giants with accurate distance estimation, with 15,454 of them classified as halo star, but we only use the 10,651 halo K giants in northern hemisphere where the whole regions are well covered by the data. We obtained a number of $6.5 \times \rm 10^{4}$ halo K giants after the correction of selection function in northern hemisphere. A factor of 2 should be multiplied considering the missing data in the southern hemisphere. With the orbital integration, we found the halo interlopers distributed at $r<5$ kpc contribute $\sim16\%$ of the whole halo stars at $r<50$ kpc, but not directly observed. We thus have to multiply a faction of 1/0.84 to obtain the K-giant number at the full region of $r<50$ kpc. At the end, we obtain the total number of halo K giants at the full region of $r<50$ kpc to be $\sim 1.6 \times \rm 10^{5}$.


As we described in Section \ref{sec:data}, in order to obtain accurate distance, we made complicated selections from the original $\sim 1.1$ million LAMOST K giants. To estimate the real number of K giants in the halo, we have to consider the ratio of halo stars in the original LAMOST K giants to our final halo sample. In the original sample, we are lacking of accurate distance estimation from \citet{Xue14}, but have a less accurate distance estimation from \citet{Carlin15}. Combining this rough distance estimation and other parameters, we calculate a rough value of Galactocentric azimuthal velocity $\tilde{\vphi}$ for all stars in the original sample. Then we take a simple cut of $\tilde{\vphi} > 0$ to select a clear subsample of halo stars, with disk stars having $\tilde{\vphi} < 0$. Our final sample are mainly in the color range of $1.0 < G_{\rm BP} - G_{\rm RP}  < 1.6$, and within this color range, the number of northern hemisphere K giants with $\tilde{\vphi} > 0$ in the original sample is 26754, while that resulting in our final sample is 4561, with a ratio of $\sim5.9$.


Within the color range of $1.0 < G_{\rm BP} - G_{\rm RP}  < 1.6$, by multiplying the factor of 5.9, we obtain the total number of halo K giants within 50 kpc to be $7.8 \times \rm 10^{5}$, with a relative difference of only $\sim2\%$ among the seven MW models. In the Galactic center region with $\rgc < 5$ kpc, the total number of halo interloping K giants is $1.2 \times 10^5$, with a relative difference of $2.5\%$ among the seven MW models.

\subsubsection{The luminosity of the stellar halo} 

\begin{figure*}
\centering
    \hspace{0.0cm}  \includegraphics[width=.99\textwidth]{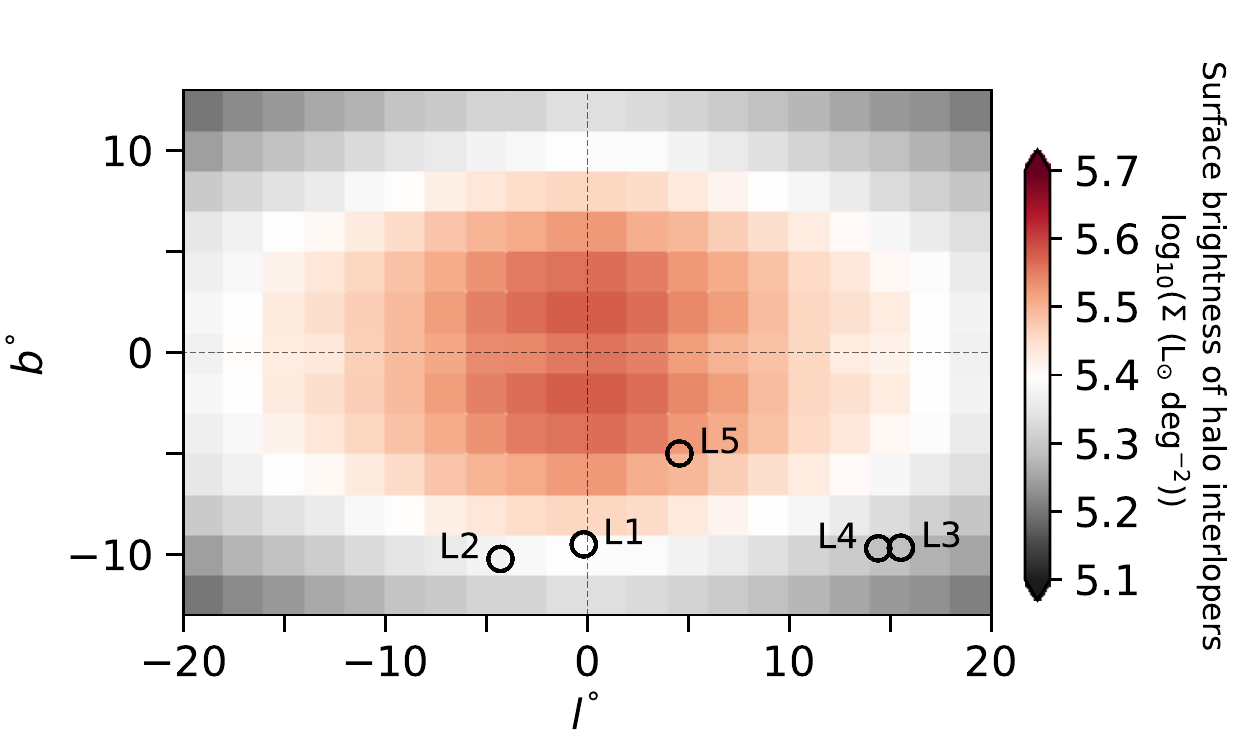}
    \hspace{-2.5cm} \includegraphics[width=1.0\textwidth]{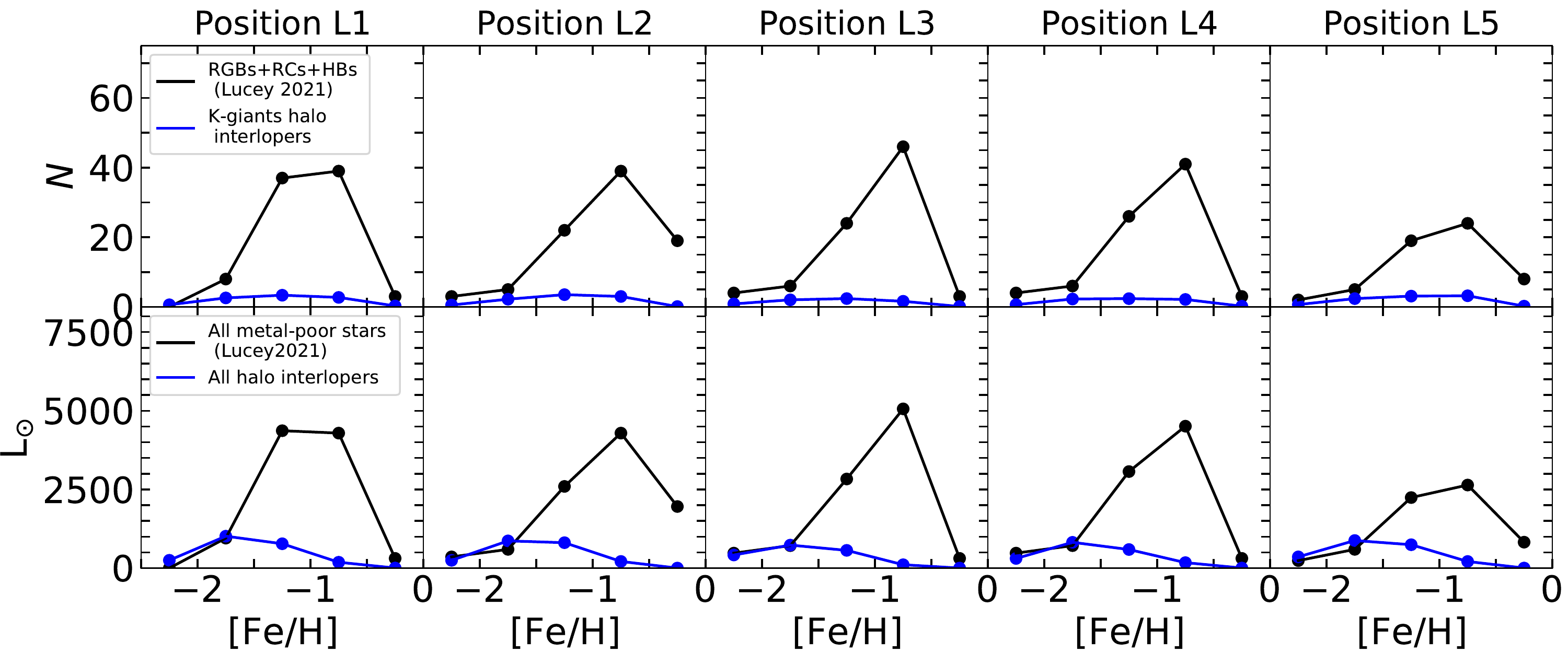}
\caption{
{\bf Top}: the surface brightness of halo interlopers toward the Galactic center derived from the particles integrated from our sample, shown in the Galactic coordinates $(l,b)$. The black circles labeled L1 to L5 indicate the location of five ARGOS target fields in \citet{Lucey21}. 
{\bf Bottom}: comparison of star number/luminosity as a function of metallicity between the halo interlopers expected from orbit integration of our sample (blue curves) and the stars directly detected in the Galactic center \citep{Lucey21} (black curves) in the five fields L1-L5.
The first row shows the number of stars directly.
The second row shows the stellar luminosity of the general halo populations normalized from the detected RGBs+RCs+HBs, and K-giant halo interlopers, respectively. The luminosity of halo interlopers is almost the same as stars directly detected with $\feh<-1.5$ in L1-L5.}
\label{halo_frac}
\end{figure*}


We then obtain the luminosity of halo stars by using the ratio of total luminosity to K giants numbers in isochrones with certain age and metallicity.
We use the isochrones from the Dartmouth Stellar Evolution Database \citep{Dotter08}, and select stars with the color range of $1.0 < G_{\rm BP} - G_{\rm RP}  < 1.6$, same as \citet{Deason19} (see Appendix \ref{app:lum} for more detail). Using a different isochrone does not affect our results. 
We assume the halo stars are old with ages from 10 to 14 Gyr, and with metallicity from $-$2.5 to 0.0. 

For each isochrone, the ratio of the total luminosity to the number of halo K giants:
\begin{equation} \label{eq:ln}
    \frac{L_{{\rm halo}, i}}{N_{{\rm KG}, i}} = 
    \frac{\int^{m_{\rm max}}_{m_{\rm min}}\ L_{i}(m)\ \xi(m)\ dm}
    {\int^{m_1}_{m_2} \xi(m)\ dm},
\end{equation}
where $i$ is the $i_{\rm th}$ isochrone with a particular age and metallicity, $\xi(m)$ is the initial mass function (IMF), and we adopt Kroupa IMF \citep{Kroupa01}, $L_{i}(m)$ is the mass-luminosity relation given by the isochrone. 
By averaging in the age range of 10-14 Gyr, we obtain ${L_{{\rm halo}, i}}/N_{{\rm KG}, i}$ = 1255, 1006, 631, 248, 51 for $\feh = [-2.5,-2], [-2,-1.5], [-1.5, -1], [-1,-0.5]$, and $[-0.5,0]$, respectively.
We obtain the overall ${L_{\rm halo}}/N_{\rm KG}$ by weighting them with the metallicity distribution of our halo K-giant sample. The final ${L_{\rm halo}}/N_{\rm KG}$ for our halo K giants sample is 666.

Within color range of $1.0 < G_{\rm BP} - G_{\rm RP}  < 1.6$, we thus obtain the total halo luminosity within 50 kpc to be $5.2 \times 10^{8}\ L_\odot$. 
In the Galactic central regions with $\rgc < 5$ kpc, we expect the total luminosity of halo interlopers within this region is $7.7 \times 10^7\ L_\odot$, and that with $\feh < -1.5$ is $3.1\times 10^7\ L_\odot$. The surface brightness of halo interlopers toward the Galactic center projected in the Galactic coordinates $(l,b)$ is shown in the top panel of Figure \ref{halo_frac}.

The stellar mass-to-light ratios are 1.3, 1.5, and 2.8 for Chabrier \citep{Chabrier03}, Kroupa, and Salpeter \citep{Salpeter55} IMF. We take the Kroupa IMF as our fiducial model, and obtain the stellar halo mass within 50 kpc to be $7.8 \times10^{8}\ M_\odot$, and the local stellar halo density of $\rho_0 = 5.8 \times 10^{-5}\ M_\odot\ \rm pc^{-3}$,
which is generally consistent with a wide range of $\rho_0 = 3$ - $15 \times 10^{-5}\ M_\odot\ \rm pc^{-3}$ allowed in the literature \citep{Morrison93, Gould98, Digby03, Juric08, de_Jong10, Bell08, Deason19}. Taken Kroupa IMF, the total mass of halo interlopers expected at $\rgc < 5$ kpc is thus $1.2 \times 10^8\ M_\odot$, and that with $\feh < -1.5$ is $4.7 \times 10^7\ M_\odot$.

\subsection{Contribution of halo interlopers to the metal-poor stars in the Galactic center} \label{sub:interlopers}



Complete observation of metal-poor stars with $\feh<-1$ in the crowded Galactic central regions is challenging.
In most surveys toward the Galactic center, such as GIBS \citep{Zoccali14}, BRAVA \citep{Kunder12}, ARGOS \citep{Freeman13}, and APOGEE \citep{Schultheis17, Zasowski19}, the Galactic center metal-rich component has been studied in detail, but the observation of metal-poor stars is not complete \citep{Arentsen20b}.
\citet{Lucey21} designed a highly efficient survey for metal-poor stars toward the Galactic center, using ARGOS spectra and SkyMapper photometry to select metal-poor giants for median- and high-resolution spectroscopic follow up. They got a sample of 595 metal-poor stars around the 25' of the five ARGOS targets and have 523 high-quality stars. The fields of five ARGOS targets are shown as black points in Galactic coordinates $(l, b)$ in the top panel of Figure \ref{halo_frac}.

In \citet{Lucey21}, the majority of stars are RGBs, RCs, and Horizontal branch stars (HBs) (see Figure 1 of \citet{Lucey21}). In the color-magnitude map, these stars are located in the range of $M_K$=(-5.0, 1.0) in 2MASS K-band magnitude and $J-K$=(0.4, 1.0). The subgiants in their sample are likely contaminated from the foreground disk along the line of sight toward the Galactic center. We will eliminate the disk stars in the later distance cut.

In order to make a direct comparison to the halo interlopers,  we convert the number of RGBs+RCs+HBs to the luminosity of metal-poor stars in the Galactic center in general ${L_{\rm GC}}$ using Equation (\ref{eq:ln}), with isochrones in the same range of age (10 - 14 Gyr) and metallicities ($\feh =-2.5$ - $0.0$) as for the halo stars. We obtain ${L_{{\rm GC}, i}}/N_{{\rm RGBs+RCs+HBs}, i}$ = 119, 119, 118, 110, 103 for $\feh = [-2.5,-2], [-2,-1.5], [-1.5, -1], [-1,-0.5]$, and $[-0.5,0]$, respectively.

In Figure \ref{halo_frac}, we compare the star number/luminosity as a function of metallicity between the halo interlopers expected from the orbit integration of our sample and the stars directly detected in the five fields L1-L5.
In each field, we take halo interlopers in the same field in $(l,b)$, and then choose the stars with $1 < \rgc < 5$ kpc and $x < 0$, the same as the location of the RGBs+RCs+HBs sample observed in \citet{Lucey21}. 

We first directly compare the number of stars as a function of metallicity in the two samples (the middle row of Figure~\ref{halo_frac}), then we compare the luminosity of halo interlopers in general to that of the metal-poor stars observed in the Galactic center (bottom row of Figure~\ref{halo_frac}). 
Taking the average of the five fields L1-L5, the luminosity fractions of halo interlopers to that of the Galactic central metal-poor stars are 100\%, 100\%, 23\%, 4\%, and 0.3\%, in the metallicity range of $[-2.5,-2.0]$, $[-2.0,-1.5]$, $[-1.5, -1.0]$, $[-1.0,-0.5]$, and $[-0.5,0]$, respectively. 
The luminosity fraction of halo interlopers to all the Galactic center metal-poor stars with $-2.5<\feh < 0$ is 23\%. The detection of stars in the Galactic center is not guaranteed to be $100\%$ complete even for the metal-poor stars with $\feh<-1$ where \citet{Lucey21} aimed to, this could be one of the major sources of uncertainty of our results. While for the relative metal-rich stars with $-1<\feh < 0$, the detection fraction in the sample of \citet{Lucey21} is low, the fractions of halo interlopers we obtained thus should be taken as upper limits. 

The detection of metal-poor stars in the Galactic center has been increased by two orders of magnitude using Gaia DR3 \citep{Rix22}. They obtained that the stellar mass at $\feh<-1.5$ within 5 kpc is $\gtrsim 5\times10^7\ M_\odot$, and could be as high as $>10^8\ M_\odot$ considering the large correction of dust obscuration, and the fraction of halo interlopers with $r_{\rm apo} < 5$ kpc is minor ($<50\%$).
While the halo interlopers within 5 kpc expected from orbit integration of our sample at $\feh < -1.5$ is $4.7 \times10^7\ M_\odot$ taken the Kroupa IMF. By comparing to the mass of $>10^8\ M_\odot$ from \citet{Rix22}, the fraction of halo interlopers is thus $\lesssim50\%$ for stars at $\feh < -1.5$ within 5 kpc, roughly consistent with the results from orbit analysis of the sample in \citet{Rix22}. However, this is different from the value of $\sim100\%$ halo interloper at $\feh < -1.5$ from the comparison with \citet{Lucey21} aforementioned. The difference is caused by the still largely uncertain mass of metal-poor stars observed in the Galactic central regions.

The fraction of halo interlopers to the Galactic center metal-poor population we obtained is generally consistent with the other results from analyzing the orbits of metal-poor stars directly detected at the Galactic center. For example, \citet{Kunder20} found that 25\% of their RRL stars with metallicity of $\sim -1.0 \pm 0.25$ in the bulge region are likely halo interlopers. \citet{Lucey21} concluded that $\sim 50\%$ of their bulge region stars with metallicity within [$-3.0$, $0.5$] could be halo interlopers, and the rate of halo interlopers decreased steadily with increasing metallicity across the full range of their sample.

\begin{figure}
    \hspace{-0.8cm} \includegraphics[width=.50\textwidth]{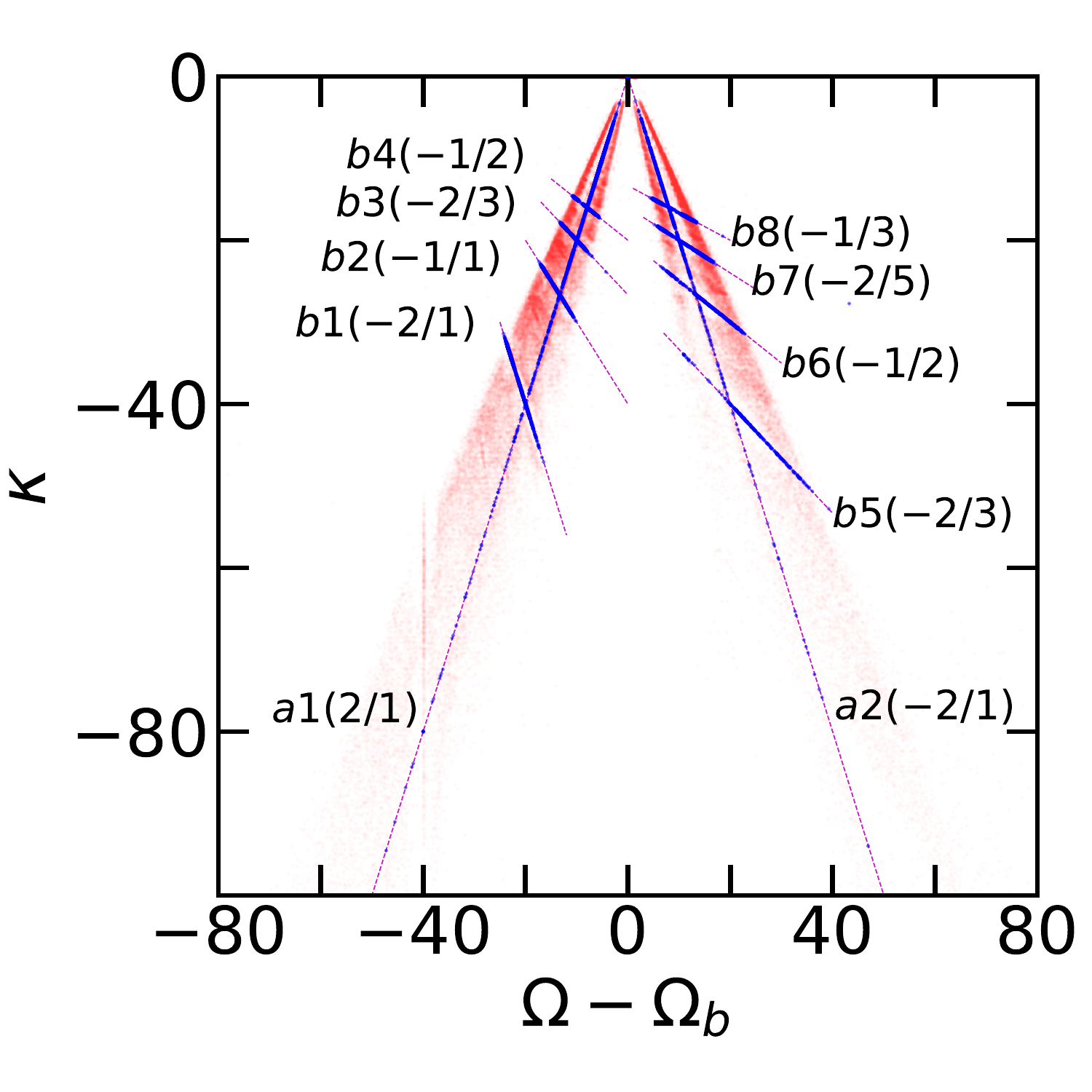}
\caption{The frequency map of $(\Omega - \Omega_b, \kappa)$ for $\sim100,000$ orbits shown as red dots (10 orbits for each star). The orbits in this map are integrated in the MW model M3. There are 10 bar-resonant orbits, shown by blue dots, labeled $a1$, $a2$, $b1$, ..., $b8$, and their slopes are indicated after the labels. }
\label{fp_fR}
\end{figure}

\begin{figure*}
    \hspace{-0.2cm} \includegraphics[width= 0.5\textwidth]{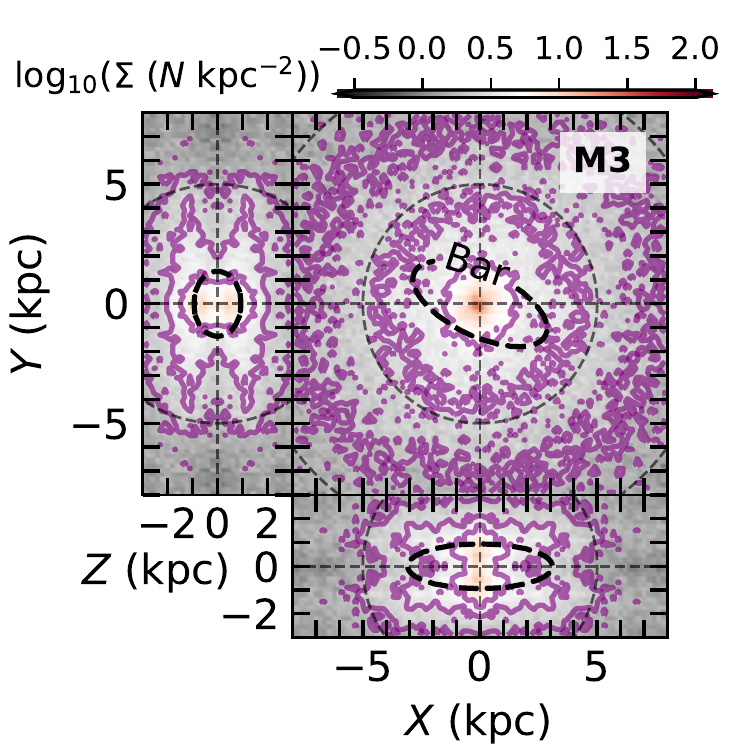}
    \hspace{-0.2cm} \includegraphics[width= 0.5\textwidth]{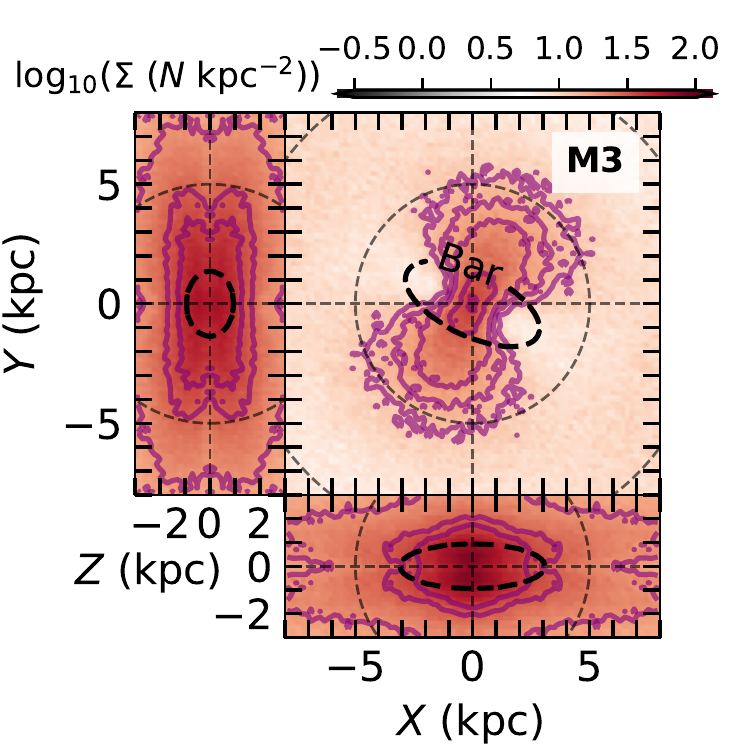}
\caption{The surface density of halo K giants on bar-resonant orbits shown in the co-rotating frame of the bar. The left and right panels display the density distributions from the $a1-a2$ sequences and the other eight sequences $b1-b8$, respectively. The bar-resonant orbits are calculated from the MW model of M3. The black dashed oval indicates the Galactic bar. There is no clear bar feature constructed by $a1-a2$ aligned with the Galactic bar, but a bar-like structure perpendicular to the Galactic bar constructed by $b1-b8$ as shown in the right panel.}
\label{od_bar}
\end{figure*}

\subsection{Halo stars on bar-resonant orbits} \label{sec:bar_resorbs}

We use orbits frequency to analyze the possible bar-trapped orbits. Orbits in the Galaxy can be described with oscillations in three directions, radial frequency $\kappa$, azimuthal frequency $\Omega$, and vertical frequency $\nu$. If the bar rotates at a pattern speed $\Omega_b$, this kind of perturbation will produce resonant orbits that satisfy $m \kappa = n (\Omega - \Omega_b)$, where $m$ and $n$ are integers, and $\Omega - \Omega_b$ is the azimuthal frequency of the resonance in the frame that corotates with the bar \citep{Molloy15, Williams16}.

We calculate the frequencies of our orbital sample using the  Numerical Analysis of Fundamental Frequencies (NAFF) code \citep{Valluri10}. The software calculates the fundamental frequencies in Cartesian and cylindrical coordinates using Fourier spectra for the phase-space coordinates of given orbits. 

In Figure \ref{fp_fR}, we present the orbital frequency map of $(\Omega - \Omega_b, \kappa)$, the stars' orbits are integrated in MW model M3. In this figure, we identify two sequences of resonant orbits with slopes of $2/1$ and $-2/1$, labeled as $a1$ and $a2$, and eight sequences that cross a1 and a2, with slopes of $-2/1,\ -1/1,\ -2/3,\ -1/2$ labeled as b1 to b4 when $\Omega - \Omega_b<$ 0, and with slopes of $-2/3,\ -1/2,\ -2/5,\ -1/3$ labeled as b5 to b8 when $\Omega - \Omega_b>$ 0.

In the two-dimensional $N$-body simulations, the bar mainly comprises stars close to stable periodic orbits families \citep{Contopoulos80}. For example, the $x_1$ family, which is the main family supporting the bar and is elongated along the bar, the $x_2$ and $x_4$ family, which have less impact on the bar and are perpendicular to the bar \citep{Contopoulos89, Sellwood93, Sellwood14}. 
All these families could be characterized by the ratio of frequency $\kappa/(\Omega - \Omega_b) = 2$ \citep{Portail15, Williams16, Perez-Villegas17}. The two sequences labeled as $a1$ and $a2$ thus should include these orbits. However, there are also a significant number of bar-resonant orbits with different frequency ratios \citep{Smirnov21}, so the other sequences we identified should belong to this category.

In Figure \ref{od_bar}, we present the surface density distribution of all the bar-resonant orbits, with stars on sequence $a1-a2$ in the left panel, and those on $b1-b8$ in the right panel. There is no clear bar feature from the $a1-a2$ sequences, but a bar-like structure perpendicular to the Galactic bar constructed by stars on $b1-b8$. These structures are slightly different from different MW models, similar figures for stars integrated in the other six MW potential models are shown in the Appendix \ref{app:bar}.

The number of the halo K giants (after correcting the selection function) on these bar-resonant orbits is $\sim 10^4$, accounting for $8\%$ of the total number of halo K giants in our orbit integrating sample. These stars are only trapped in the bar-resonant orbits with a short time scale, their contribution in the inner 5 kpc is only $\sim0.8\%$ of the total number of halo interlopers. The metallicity of these bar-resonant halo K giants is not significantly different from the whole halo K-giant sample, with the peak of metallicity distribution at $\feh\sim-1.4$.

\section{Discussion} \label{sec:Dis}

We selected halo stars combing the distribution in metallicity and Galactocentric azimuthal velocity $\vphi$. Stars from the controversial ``Splash" structure \citep{Belokurov20, Amarante20} are included in our sample. These stars are supposed to be born in-situ in the disk but are heated by ancient mergers and are currently a halo-like structure with $\feh >$ -0.7 \citep{Belokurov20}. In this section, we discuss the impact of the Splash by removing the stars with $\feh > -1$ from our sample of halo K giants.

After removing the stars with $\feh > -1$, we find that the halo rotation becomes faster at $r<5$ kpc with $v_\phi$ increasing from $\sim -10\,\kms$ (Figure~\ref{rvl}) to $\sim-20$ $\kms$ and the halo becomes more spherical at $r<5$ kpc with $q$ increasing from $\sim 0.5$ (Figure~\ref{ma_rq_nu}) to 0.8. This is consistent with the Splash stars being on highly eccentric orbits and with little to no angular momentum \citep{Belokurov20}. 
 At the same time, the metallicity gradient downs to $\sim -0.022$ dex kpc$^{-1}$ at inner region ($\rgc < 5$ kpc) and downs to $\sim -0.002$ dex kpc$^{-1}$ at outer region ($\rgc > 5$ kpc).
The Splash structure seems to be one of the major origins of the metallicity gradients in the stellar halo.

After removing these stars, the number of the halo K giants on the bar-resonant orbits will decrease by a fraction of $10-15\%$ based on different MW potential models. This is consistent with the fraction of stars with $\feh>-1$ at $r<10$ kpc in our sample.

\section{Conclusion}
\label{sec:sum}
We construct the full stellar density distribution of the MW halo at $\rgc<50$ kpc using K giants cross-matched from LAMOST DR5 and Gaia DR3, with 6d position-velocity information and metallicity measured. The density distribution is constructed by carefully considering the selection function and integrating the stellar orbits of halo stars in fixed MW potentials. The major results are as follows:
\begin{itemize}

\item {The stellar halo's density distribution can be described by a double-broken power-law function with power-law coefficients $\alpha = -1.5, -2.8, -6.1$ at $r_{\rm ellip} < 10$ kpc,  $10< r_{\rm ellip} < 25$ kpc, and $r_{\rm ellip} >25$ kpc, respectively. This density profile is similar to the GSE simulation of \citet{Naidu21}, which implies that GSE members might be the dominant stars in the inner halo. The stellar halo is near-spherical at $\rgc>$ 30 kpc with $q\sim 0.8$ and becomes flattered in the inner regions, it has $q\sim 0.5$ at $\rgc<$ 5 kpc.}

\item {The stellar halo is highly radially anisotropic ($\beta\sim0.8$) with mild retrograde rotation of $10\ \kms$ at outer regions, while it becomes isotropic with a small prograde rotation of $-10$ to $-20\ \kms$ in the inner 10 kpc. The velocity dispersions of stellar halo reach $\sim 250\ \kms$ in the Galactic center.}

\item {The metallicity distribution from $\sim5$ to $50$ kpc is shallower with a gradient of $-0.005$ dex kpc$^{-1}$, and becomes steeper at $\rgc<$ 5 kpc with a gradient of $\sim -0.029$ dex kpc$^{-1}$. The peak of metallicity distribution is at around $-1.3$, there is an excess of relative metal-rich stars with $\feh>-1$ in the inner 10 kpc.}

\item {We obtain the total stellar halo mass within 50 kpc to be $\sim 7.8\times10^8\ M_{\odot}$, and the mass within 5 kpc to be $\sim 1.2\times10^8\ M_{\odot}$. By comparing with the number of metal-poor stars directly observed in the Galactic center, we find the fraction of halo interlopers is $23\%$ for stars with $-2.5<\feh<0$ in the Galactic center, the fraction of halo interlopers increases with decreasing metallicity, and it is $50 - 100\%$ for stars with $\feh<-1.5$.}

\item {By analyzing the orbit frequency, we find $\sim8\%$ of halo K giants are temporarily trapped on bar-resonant orbits. But the halo K giants on bar-resonant orbits only contribute to $\sim0.8\%$ of the total halo interlopers at $r<5$ kpc, and there is not a clear bar-like structure.}
\end{itemize}

\vspace{5mm}




We thank the discussion with Ortwin Gerhard.
L.Z. acknowledges the support from the National Key
R$\&$D Program of China under grant No. 2018YFA0404501, National Natural Science Foundation of China under grant No. Y945271001 and CAS Project for Young Scientists in Basic Research under grant No. YSBR-062.
C.Q.Y acknowledges support from the LAMOST FELLOWSHIP fund.
The LAMOST FELLOWSHIP is supported by Special Funding for Advanced Users, budgeted and administered by the Center for Astronomical Mega-science, Chinese Academy of Sciences (CAMS-CAS). This work is supported by the Cultivation Project for LAMOST Scientific Payoff and Research Achievement of CAMS-CAS.
Guoshoujing Telescope (the Large Sky Area Multi-Object Fiber Spectroscopic Telescope LAMOST) is a National Major Scientific Project built by the Chinese Academy of Sciences. Funding for the project has been provided by the National Development and Reform Commission. LAMOST is operated and managed by the National Astronomical Observatories, Chinese Academy of Sciences.
This work has made use of data from the European Space Agency (ESA) mission Gaia (\url{https://www.cosmos.esa.int/gaia}), processed by the Gaia Data Processing and Analysis Consortium (DPAC, \url{https://www.cosmos.esa.int/web/gaia/dpac/consortium}). Funding for the DPAC has been provided by national institutions, in particular the institutions participating in the Gaia Multilateral Agreement.
\appendix

\section{A comparison with StarHorse's distance} \label{app:dist}
\restartappendixnumbering

\begin{figure}
\centering
    \hspace{0.0cm} \includegraphics[width=.40\textwidth]{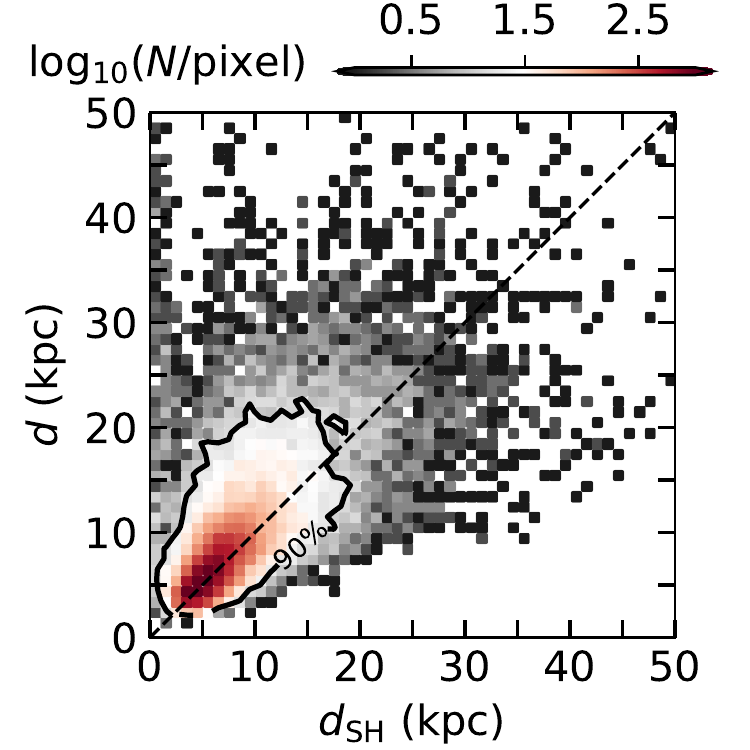} \\ 
\caption{The comparison between our distance and {\tt StarHorse} distance $d_{\rm SH}$, the dashed line indicates a 1:1 ratio, and the black contour includes 90\% of the stars. {\tt StarHorse} failed to give a proper estimation for a large fraction of stars at $d>20$ kpc, for which they just give zero values.}
\label{A.starhorse}
\end{figure}

Figure \ref{A.starhorse} displays a comparison between our distance and the distance of {\tt StarHorse} ($d_{\rm SH}$) \citep{Anders22}. The StarHorse data has been selected with SH\_GAIAFLAG = “000” and SH\_OUTFLAG = “00000”. Both distances agree with each other up to $\sim20$ kpc, but over 20 kpc, the distance provided by {\tt StarHorse} is skewed to smaller values than our distance, and they failed to give proper estimation for a large fraction of stars at $d>20$ kpc.

\section{Luminosity of stellar halo} \label{app:lum}

\begin{figure}
\centering
    \hspace{0.0cm} \includegraphics[width=.40\textwidth]{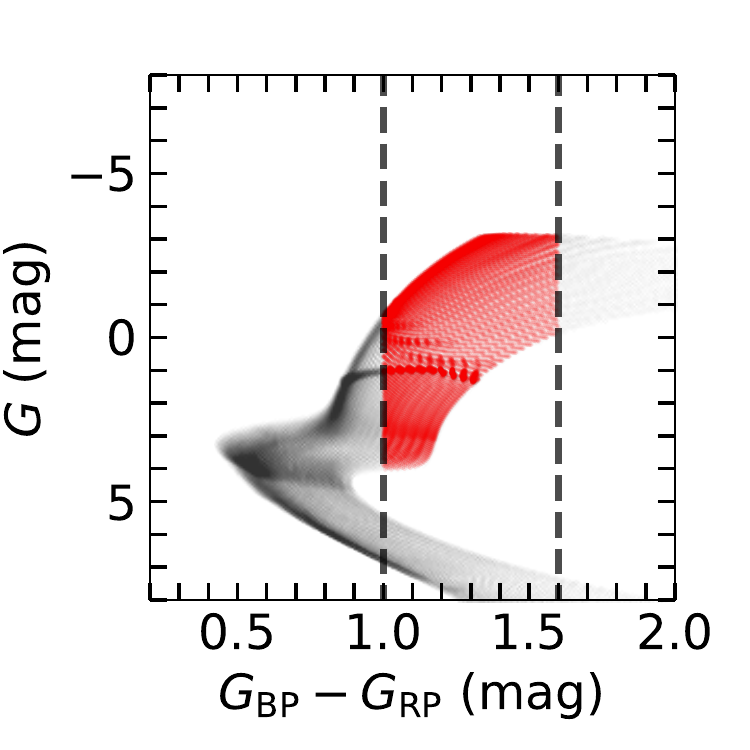}
    \hspace{0.0cm} \includegraphics[width=.40\textwidth]{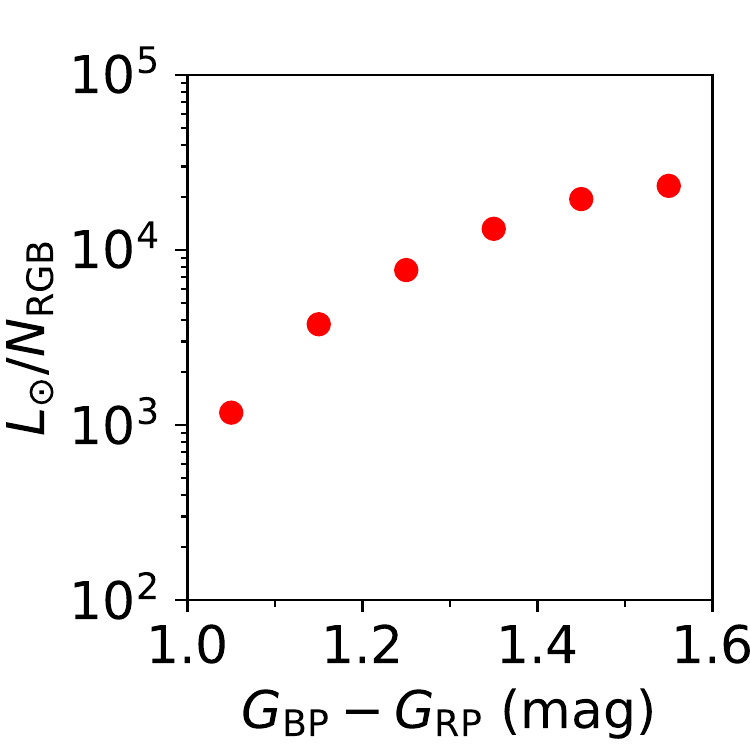} \\ 
\caption{
{\bf Left}: absolute magnitude versus color for stars from Dartmouth isochrones. Giant stars with color range of $1.0 < G_{\rm BP} - G_{\rm RP}  < 1.6$ are shown in red.
{\bf Right}: The relation between total luminosity and number of RGB stars per 0.1 color bin.
}
\label{B.lum}
\end{figure}

In Figure \ref{B.lum}, we show the selected halo isochrones from the Dartmouth Stellar Evolution Database \citep{Dotter08} and the relation between luminosity and the number of RGBs. The ${L_{\rm halo}}/N_{\rm KG}$ in each 0.1 bin obtained from Dartmouth isochrone is in good agreement with that obtained by \citet{Deason19} from PARSEC isochrones \citep{Bressan12}.

\section{Bar-resonant orbits} \label{app:bar}

\begin{figure}
    \hspace{-0.2cm} \includegraphics[width= 0.25\textwidth]{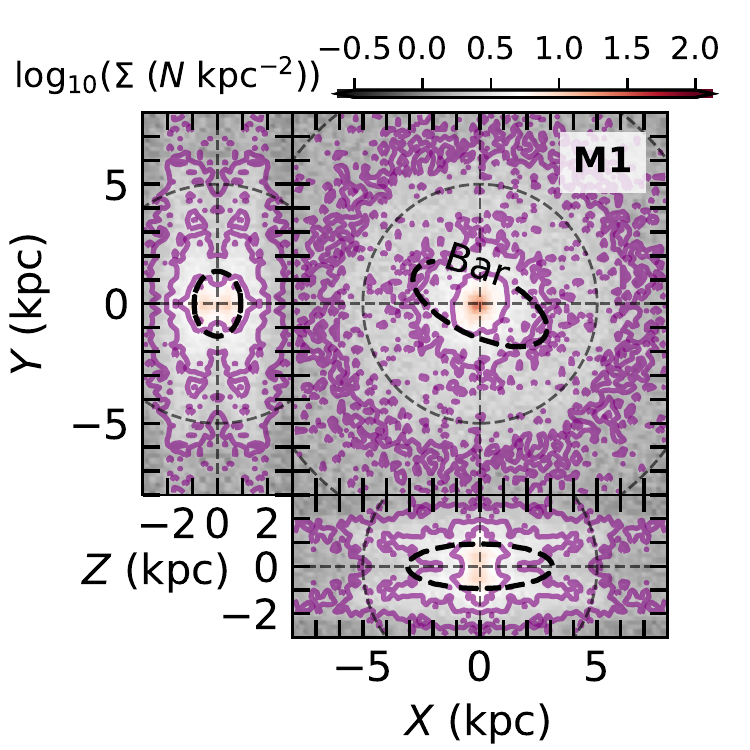}
    \hspace{-0.2cm} \includegraphics[width= 0.25\textwidth]{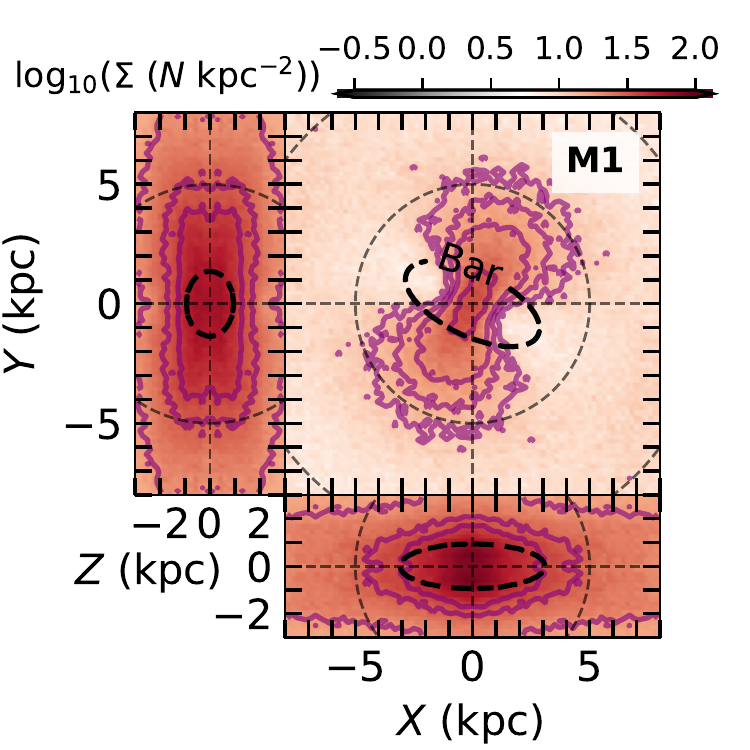}
    \hspace{-0.2cm} \includegraphics[width= 0.25\textwidth]{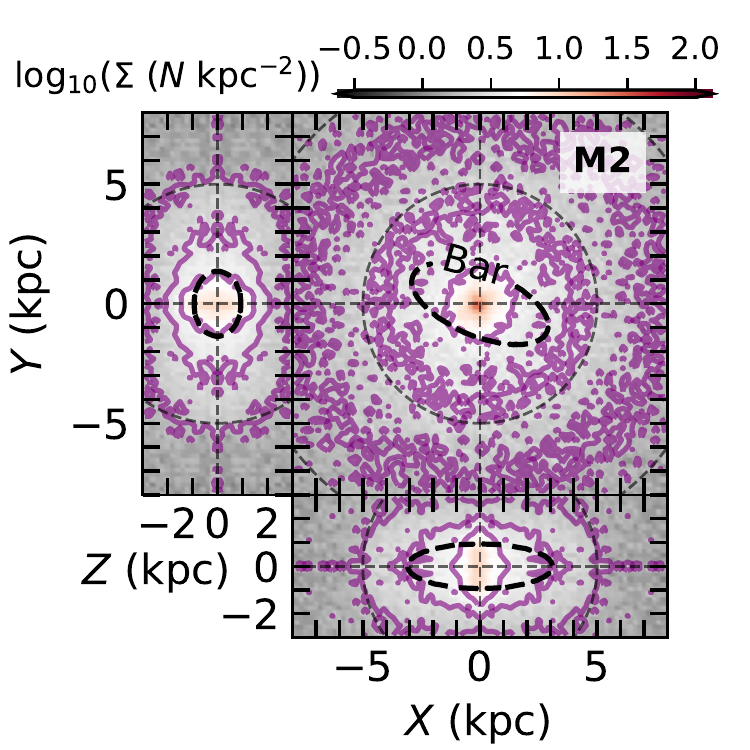}
    \hspace{-0.2cm} \includegraphics[width= 0.25\textwidth]{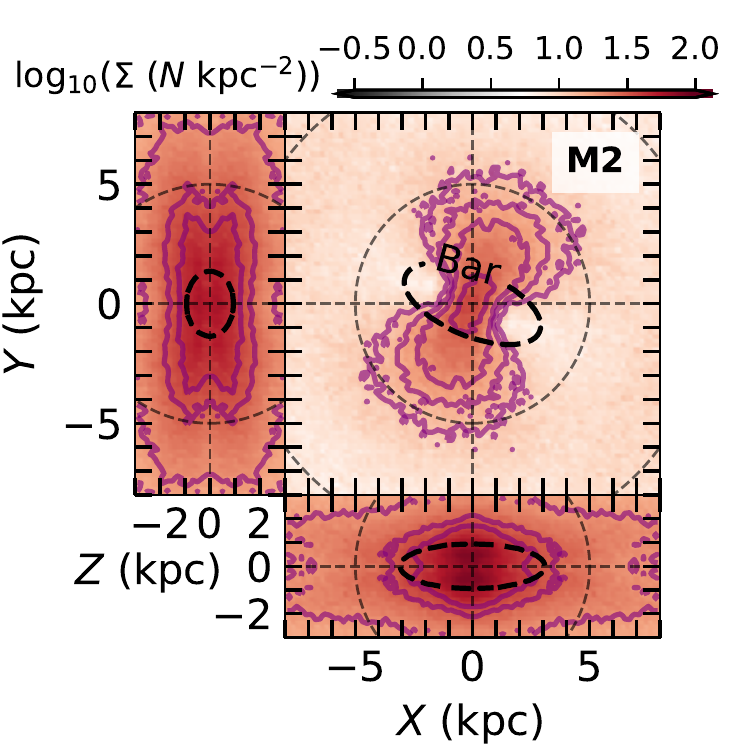}\\
    \hspace{-0.2cm} \includegraphics[width= 0.25\textwidth]{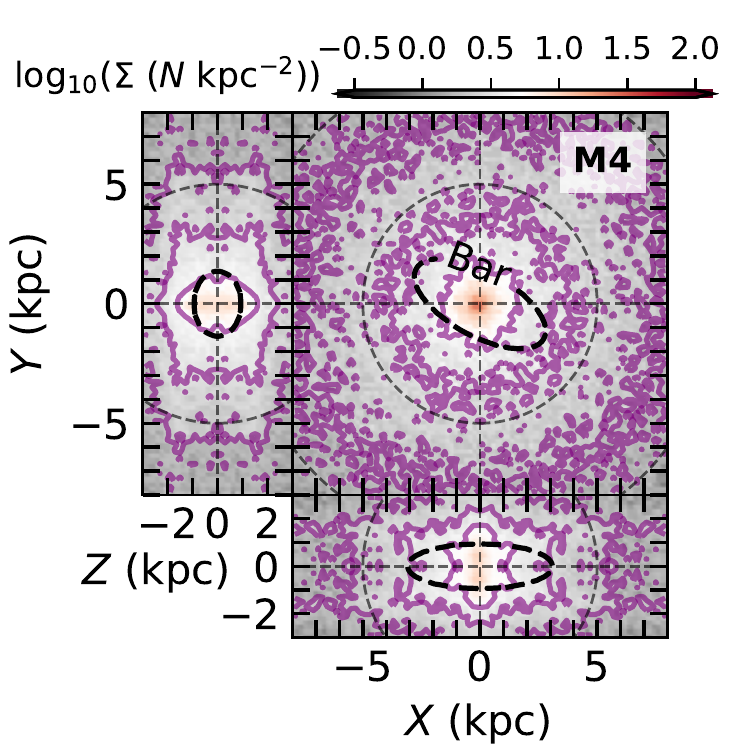}
    \hspace{-0.2cm} \includegraphics[width= 0.25\textwidth]{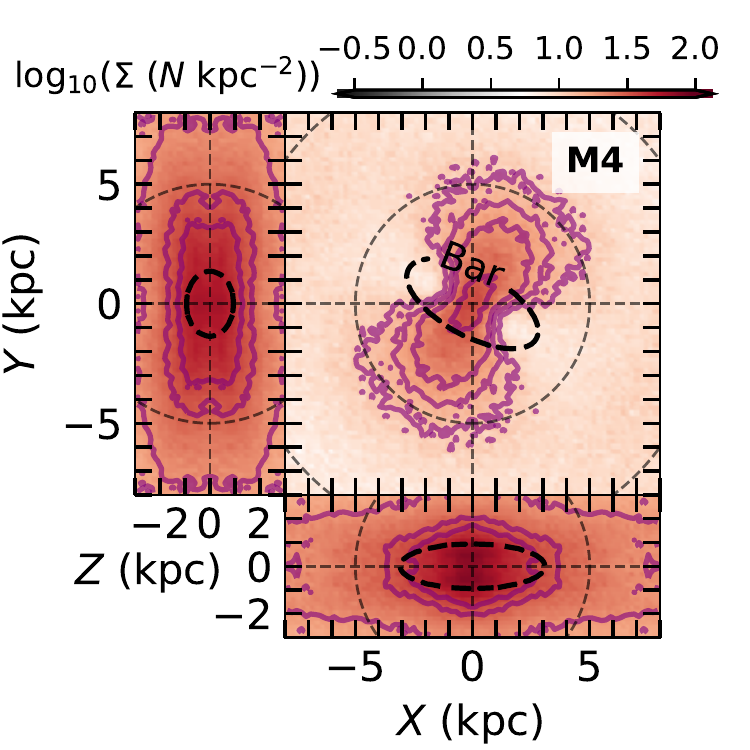}
    \hspace{-0.2cm} \includegraphics[width= 0.25\textwidth]{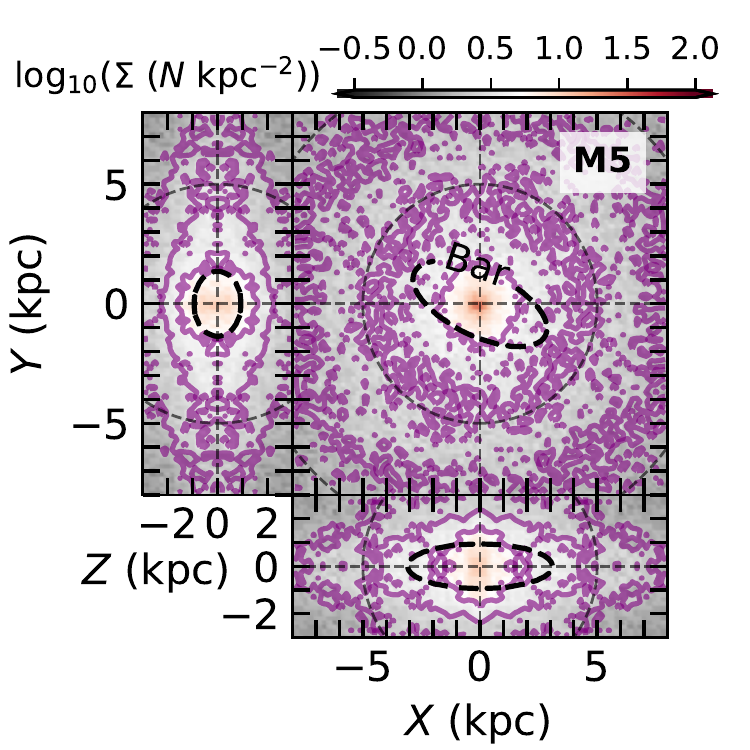}
    \hspace{-0.2cm} \includegraphics[width= 0.25\textwidth]{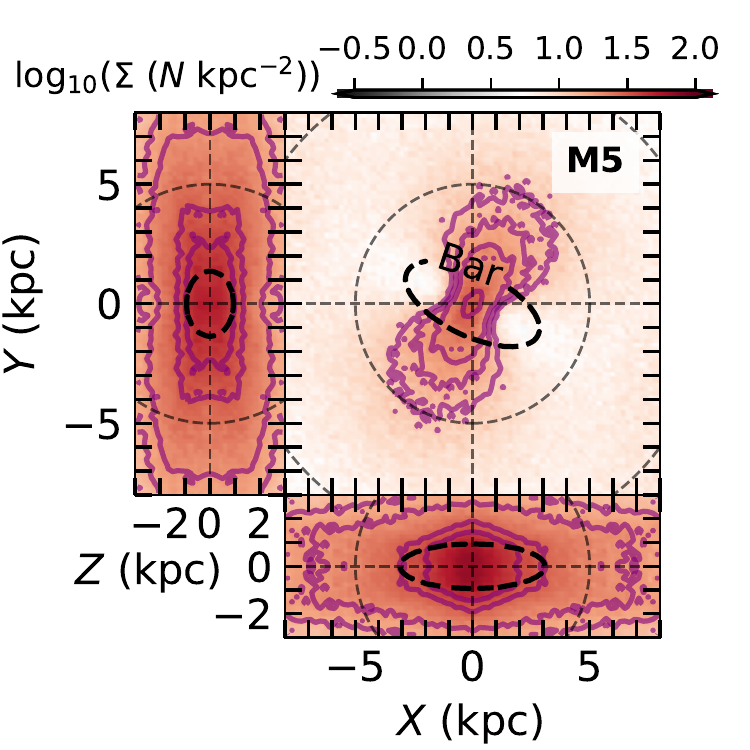}\\
    \hspace{-0.2cm} \includegraphics[width= 0.25\textwidth]{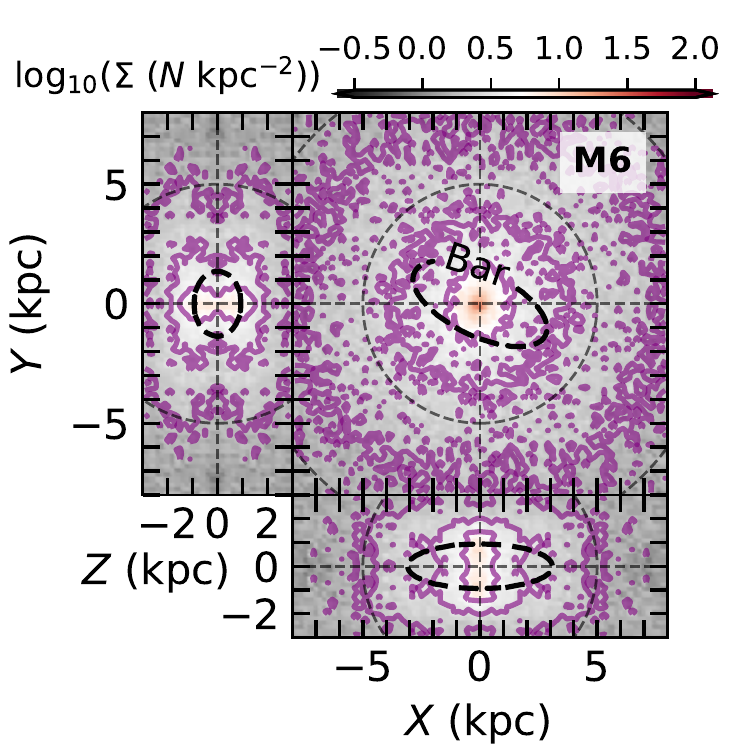}
    \hspace{-0.2cm} \includegraphics[width= 0.25\textwidth]{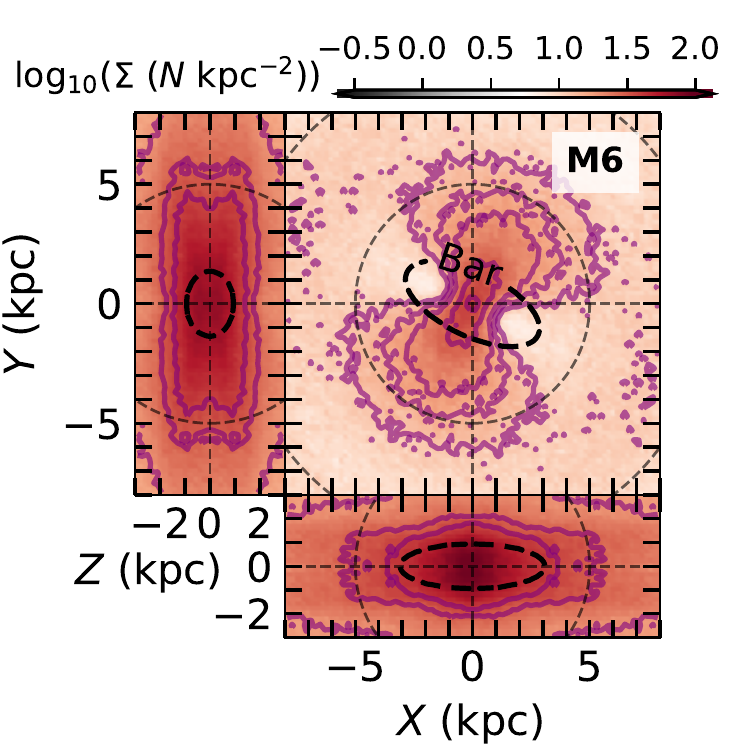}
    \hspace{-0.2cm} \includegraphics[width= 0.25\textwidth]{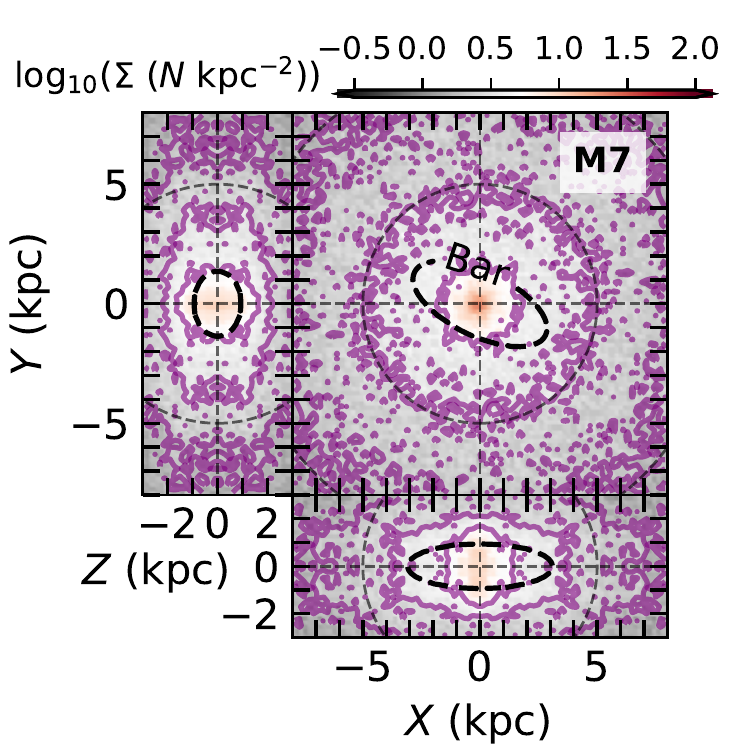}
    \hspace{-0.2cm} \includegraphics[width= 0.25\textwidth]{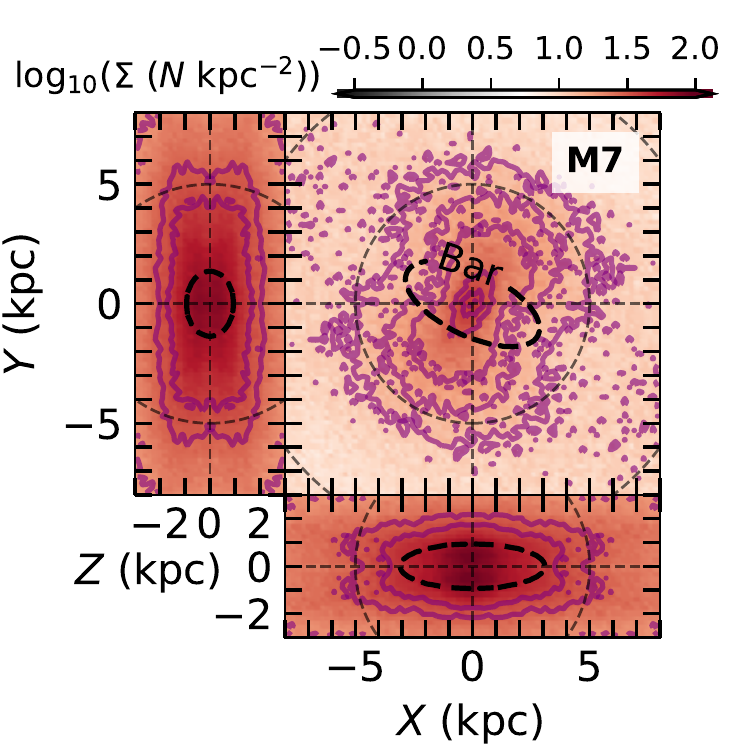}
\caption{The surface density of halo K giants on the bar-resonant orbits shown in the corotating frame of the bar. The orbits are calculated in six different MW models as labeled in the upper right corner (M1, M2, M4, M5, M6, M7).}
\label{C.vbar_omg_in}
\end{figure}

Figure \ref{C.vbar_omg_in} shows that bar-resonant orbits' surface density differs from different MW models. With the increase of the pattern speed (M5 to M6) or the increase of halo mass (M1 to M7), the bar-like feature becomes prominent, which means more bar-resonant stars contribute to the inner region, and there is no apparent difference between different bar angles (M2 to M4).

\clearpage

\bibliography{1.bibtex}{}
\bibliographystyle{aasjournal}

\end{document}